\input{epsf}

\documentclass[12pt, draftclsnofoot, peerreview, onecolumn]{IEEEtran}
\usepackage{epsf}
\usepackage{graphicx}
\usepackage[cmex10]{amsmath}
\usepackage{amsthm}
\usepackage{amssymb}
\usepackage{epsfig,latexsym,amsmath,epsf,amssymb,amsfonts}
\usepackage{algorithm, algorithmic}
\usepackage{placeins}
\usepackage{cite}
\usepackage{comment}
\usepackage{subfigure}
\usepackage{lipsum}
\usepackage{color}

\usepackage{bbm}

\usepackage{pgf,tikz}
\usetikzlibrary{shapes,arrows,automata, calc}

\newtheorem{Definition}{\hskip 0pt Definition}
\newtheorem{Theorem}{\hskip 0pt Theorem}
\newtheorem{Lemma}{\hskip 0pt Lemma}
\newtheorem{Corollary}{\hskip 0pt Corollary}

\newtheorem*{Remark}{\hskip 0pt Remark}

\begin{document}
%
\title{Stackelberg Game for Distributed Time Scheduling in RF-Powered Backscatter Cognitive Radio Networks\vspace*{-4mm}}

\author{
\IEEEauthorblockN{Wenbo Wang,
Dinh Thai Hoang,
Dusit Niyato,~\IEEEmembership{Fellow,~IEEE,}
Ping Wang,~\IEEEmembership{Senior Member,~IEEE} and
Dong In Kim,~\IEEEmembership{Senior Member,~IEEE}}
\vspace*{-6mm}}

\maketitle
\vspace*{-4mm}
\begin{abstract}
\vspace*{-4mm}
In this paper, we study the transmission strategy adaptation problem in an RF-powered cognitive radio network, in which hybrid secondary users are able to switch between the harvest-then-transmit mode and the ambient backscatter mode for their communication with the secondary gateway. In the network, a monetary incentive is introduced for managing the interference caused by the secondary transmission with imperfect channel sensing. The sensing-pricing-transmitting process of the secondary gateway and the transmitters is modeled as a single-leader-multi-follower Stackelberg game. Furthermore, the follower sub-game among the secondary transmitters is modeled as a generalized Nash equilibrium problem with shared constraints. Based on our theoretical discoveries regarding the properties of equilibria in the follower sub-game and the Stackelberg game, we propose a distributed, iterative strategy searching scheme that guarantees the convergence to the Stackelberg equilibrium. The numerical simulations show that the proposed hybrid transmission scheme always outperforms the schemes with fixed transmission modes. Furthermore, the simulations reveal that the adopted hybrid scheme is able to achieve a higher throughput than the sum of the throughput obtained from the schemes with fixed transmission modes.
\end{abstract}

\section{Introduction}
\label{sec_intro}
Thanks to the development in Cognitive Radio (CR) technologies, recently Dynamic Spectrum Access (DSA) has seen tremendous advancements in improving the efficiency of Radio Frequency (RF) spectrum management~\cite{6365154}. With the emphasis on policy and spectrum agility, DSA is also envisaged to provide more flexible business models for spectrum sharing with the non-legitimated (secondary) spectrum users. Meanwhile, with the proliferation of low-power networks such as Internet of Things (IoT), green powered CR networks with the capabilities of energy harvesting have also drawn the focus of research in addition to the studies on RF efficiency~\cite{6365154}. By enabling RF energy harvesting on the CR devices, the green powered CR network is able to opportunistically harness the free energy from primary signals as well as exploiting the underutilized spectrum. However, in a typical RF-powered CR network, the transmission is usually organized in a harvest-then-transmit manner~\cite{7353134, 6678102}. As a result, the performance of the secondary transmission is mainly dependent on the activity of the Primary Transmitters (PTs). In particular, to guarantee a satisfying performance of the RF-powered Secondary Transmitters (STs), the activity of the PTs is expected to be kept at a mild level. Otherwise, a PT with high frequent data transmission will leave the STs little time for transmitting over the idle channel, while a PT with low frequent data transmission will result in shortage of the harvested energy. In both situations, the total transmitted bits by the STs may be significantly reduced.

To overcome the performance degradation due to the uncontrollable PT activities, ambient backscattering~\cite{Liu:2013:ABW:2486001.2486015} has recently been introduced into the RF-powered CR networks~\cite{7937935}. With ambient backscattering, an ST uses modulated backscattering of the ambient signals (i.e., the primary signals), such as UHF TV or Wi-Fi signals, to communicate with the Secondary Receiver (SR). A backscattering ST can transmit passively by switching between reflecting and non-reflecting states at a much lower rate than that of the ambient signals. The target SR decodes the information from the received signal using a simple averaging mechanism~\cite{Liu:2013:ABW:2486001.2486015}. However, although the circuit power consumption of backscattering is negligible, to maintain the power level of the backscattered signal, it is impractical for an ST to harvest the RF energy while operating in the backscattering mode. Therefore, when integrating the ambient backscattering module into the existing RF-powered CR devices, a natural question arises on how to properly allocate the time resource between the two modes of backscattering and harvesting, such that the total transmitted bits of the STs are maximized.

In this paper, we consider a multi-user CR network, where the STs are able to transmit data by using the overlaid, harvest-then-transmit mode and the ambient backscatter mode in a hybrid scheme. To answer the above question on optimal resource allocation, we formulate the joint transmit-mode selection and time resource allocation problem as a constrained non-cooperative game among the STs. In addition, to cope with the interference caused by STs due to imperfect channel sensing, we introduce a pricing mechanism for the primary network to control the STs' transmit behaviors indirectly with monetary incentives. Furthermore, the resource allocation based on dynamic interference pricing is formulated as a two-stage Stackelberg game. Based on the analysis of the game properties, we propose a distributed, iterative allocation strategy searching mechanism which is guaranteed to converge to the Stackelberg Equilibrium (SE).\vspace*{-3.5mm}

\subsection{Related Work}
\label{sec_review}
\subsubsection{Resource Allocation in RF-Powered CR Networks}
\label{sub_sec_RF}
In the past few years, a great amount of effort has been put in the study of the techniques for RF-powered CR networks (see~\cite{6951347} and the references therein). In an RF-powered CR network, energy harvesting and spectrum access are performed opportunistically following a ``sensing-harvesting-transmission'' paradigm. In particular, in the scenario where the PT randomly occupies and evacuates the channel, an ST is subject to the constraints on the interference probability with the PTs as well as the constraint on the total consumable energy for transmission. Therefore, for an ST, a balance is expected to be stroke between the time allocated for sensing and the time allocated for energy harvesting. In~\cite{6783667, 6661321}, the problems of optimal pairing for the sensing duration and detection threshold in an RF-powered network are addressed through the formulation of constrained (stochastic) nonlinear programming problems. In~\cite{7009968}, the study is further extended to the scenario of cooperative sensing with data/decision, where a single ST uses the imperfect detection results from multiple mini-sensing slots to determine its operation mode. Furthermore, by extending the degree of freedom in the decision or resource variable space, the trade-off problems between sensing, harvesting and throughput have also been incorporated into the scenarios of Time-Division Multiplexing Access (TDMA) in multi-user CR networks~\cite{7866871}, multi-channel selection~\cite{6985740} and cognitive relay networks~\cite{7342973}.

\subsubsection{Ambient Backscatter and its Application in CR Networks}
Compared with the traditional backscatter devices which rely on a dedicated carrier emitter, ambient backscattering devices leverages the uncontrollable, pre-existing (i.e., ambient) RF signals for its own transmission~\cite{Liu:2013:ABW:2486001.2486015}. Since the introduction of ambient backscattering technique~\cite{Liu:2013:ABW:2486001.2486015}, considerable effort has been devoted to improving its performance in terms of transmit range, throughput and Bit Error Rate (BER)~\cite{Bharadia:2015:BHT:2785956.2787490, 7551180, 7873313}. In~\cite{Bharadia:2015:BHT:2785956.2787490}, the backscatter devices use a Wi-Fi Access Point (AP) as the ambient RF source as well as the receiver. Since the AP works as both the RF source and the receiver, and therefore knows the original signal, it is possible for the backscattering transmitter to adopt phase modulation and for the AP to implement self-interference cancellation based on standard channel estimation in the system. With the improved backscatter coding/decoding mechanism, the proposed mechanism in~\cite{Bharadia:2015:BHT:2785956.2787490} achieves a throughput of $1$Mbps at the transmit range of $5$m. Alternatively, when unknown ambient signals are used for backscattering, differential encoding and on-off keying is usually adopted at the transmitter~\cite{7551180}. Correspondingly, energy detection based on hypothesis test is usually used by the receiver/reader for decoding without the need of knowing the channel state information. In~\cite{7873313}, by allowing the transmitter's antenna to change its impedance and backscatter with 3 states, the constellation density is expanded to a ternary code from on-off keying. Thereby, a significant increase in the transmit bitrate can be achieved with the same energy detection-based decoding mechanism at the receiver.

With the convenience of requiring no dedicated infrastructure to generate carrier signals and utilizing existing ambient signals for transmission, ambient backscattering is considered especially appropriate to be incorporated into RF-powered CR networks. In addition, the functionalities required by ambient backscatter such as carrier sensing and distributed Multiple Access Control (MAC) protocol~\cite{Liu:2013:ABW:2486001.2486015} are ready-to-access in CR networks~\cite{6365154}. Meanwhile, since backscatter only creates additional paths from the backscattering transmitter to the primary receiver of the ambient signals, it can be effectively removed by the existing techniques such as multi-path distortion equalizer at the primary receiver or precoding at the PT. Also, a backscatter transmitter can offset the carrier phase by a certain frequency to avoid the interference~\cite{Kellogg:2017:PWB:3036699.3036711}. Therefore, backscatter-induced interference to the primary transmission is generally negligible~\cite{Liu:2013:ABW:2486001.2486015}. Recently, emerging applications of ambient backscatter in RF-powered CR networks have been proposed in~\cite{7937935, 7997476, 7869352}. In these works, the research focus is mostly placed on optimal operation scheduling for a single ST~\cite{7937935, 7997476}, or centralized scheduling for multiple STs~\cite{7869352}.\vspace*{-3.5mm}

\subsection{Contributions and Paper Organization}
In this paper, we study a CR network with multiple, hybrid STs that are jointly powered by RF-energy harvesting and backscattering techniques. Compared with the existing studies, we consider the practical situation for the CR network to have imperfect channel sensing capabilities, and emphasize the distributed nature of the CR network. To compensate the potential interference caused by RF-powered ST transmission, we introduce a pricing mechanism for the primary network to guide the time resource allocation among the STs. We model the interaction between the primary network and the secondary network as a single-leader-multi-follower hierarchical game. By providing a series of theoretical analyses on the properties of the equilibria in the game, we propose an iterative, distributed strategy searching mechanism that guarantees the convergence to the Stackelberg Equilibria (SE) as well as the social optimality among the STs.

The rest of the paper is organized as follows. Section~\ref{sec_system_model} describes TDMA-based sensing and transmission mechanism with joint energy-harvesting and ambient backscattering. Section~\ref{sec_stackelberg} introduces the pricing mechanism for interference compensation and proposes the Stackelberg game-based formulation of the interaction between the primary and secondary networks. Section~\ref{sec_SE_searching_method} proposes the distributed equilibrium searching method based on the game analysis presented in Section~\ref{sec_stackelberg}. Section~\ref{sec_simulation} provides the numerical simulations for performance evaluation of the proposed algorithm and Section~\ref{sec_conclusion} concludes the paper with a summary of the contribution.\vspace*{-2mm}

\section{System Model}
\label{sec_system_model}
\begin{figure}[t]
  \centering
  \subfigure[]{\label{fig_network_model}\includegraphics[width=.23\linewidth]{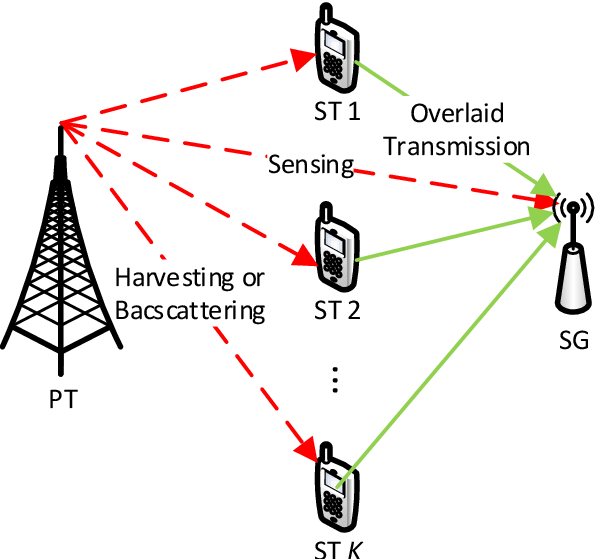}}  \hspace{5mm}
  \subfigure[]{\label{fig_sensing_slots}\includegraphics[width=.45\linewidth]{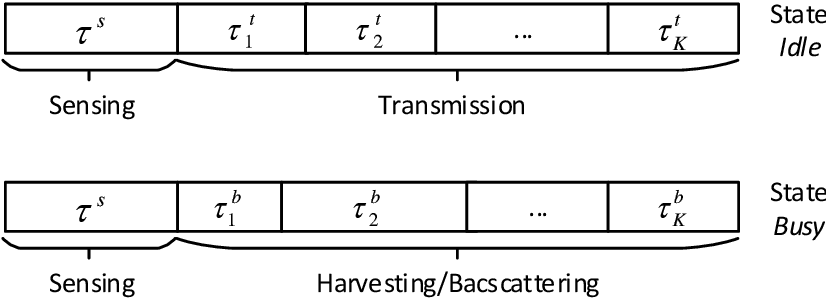}}
\caption{System model. (a) Network structure. (b) An example of sub-time slot allocation in the two channel states.}\label{Fig_illustration}
\vspace*{-2mm}
\end{figure}

We consider a multi-transmitter CR network where $K$ STs are equipped with both an energy harvesting module and a backscatter circuit (see Figure~\ref{fig_network_model}).
We assume that the PT's channel occupancy process can be modeled as a discrete-time 0-1 renewal process, where ``0'' represents the state \emph{Idle} and ``1'' represents the state \emph{Busy}. The minimum waiting time between two successive state renewal is $T$, which is also the time length of the STs' time slot. We assume that during one time slot the probabilities for the channel to be at the two states, i.e., \emph{Idle} and \emph{Busy} are known a-priori as $p_0$ and $p_1=1-p_0$, respectively. For the discrete-time renewal process, they are also the steady-state probabilities.
The STs operate in TDMA mode and transmit to the same Secondary Gateway (SG). An ST switches between the harvest-then-transmit mode and the backscattering mode for its own data transmission. When operating as an active transmitter, the ST is expected to transmit in an overlaying mode. When the channel is occupied by the primary transmission, the ST cannot transmit but is able to either harvest energy from the PT's signal or backscatter the PT's signal for its own data transmission with a relatively lower bitrate. We consider that the SG deploys an energy detector and is responsible for notifying the STs about the state of the primary channel. We assume that the secondary transmission is executed in time slots, and each time slot can be further divided into three sub-phases for channel sensing, energy-harvesting/backscattering and active transmission, respectively (see Figure~\ref{fig_sensing_slots}).\vspace*{-5mm}

\subsection{Spectrum Sensing}
Since both the harvest-then-transmit mode and the backscattering mode are of low energy, we assume that the STs are placed not too far from the SG. Therefore, the PT activities can be considered identical across the CR network. We consider that the SG is able to dynamically set up the length of the sensing phase and the detection threshold. Based on the standard detection theoretic formulation~\cite{4489760, 6253200}, for the received primary signal $y(t)$ at the SG, a binary hypothesis testing can be formulated as follows:
\begin{eqnarray}
 \label{eq:hypothesis_testing}
  y(t) = \left\{
 \begin{array}{ll}
  w(t) & : \mathcal{H}_0,\\
  \sqrt{h^{\textrm{PT}}}x(t) + w(t) & : \mathcal{H}_1,
 \end{array}  \right.
\end{eqnarray}
where $w(t)$ is the Additive White Gaussian Noise (AWGN) with the variance $\xi^2$, and $\sqrt{h^{\textrm{PT}}}x(t)$ is the received primary signal with the average power gain $h^{\textrm{PT}}$. $\mathcal{H}_0$ denotes the hypothesis that the primary channel is in state \emph{Idle}, and $\mathcal{H}_1$ denotes the hypothesis that the primary signal is in state \emph{Busy}. The performance of energy detection is measured using the sample statistics $Y=\sum_{i=1}^{N}y^2(i)$ in terms of the probabilities of false alarm $p^f=\Pr(Y>\epsilon|\mathcal{H}_0)$ and detection $p^d=\Pr(Y>\epsilon|\mathcal{H}_1)$ with the detection threshold $\epsilon$. Under the standard assumptions in the literature of channel sensing techniques in cognitive radio networks~\cite{4489760, 6253200}, $Y$ can be approximated as a Gaussian random variable, the mean and variance of which under $\mathcal{H}_0$ and $\mathcal{H}_1$ are $E(Y|\mathcal{H}_0)=\xi^2$, $E(Y|\mathcal{H}_1)=(\gamma+1)\xi^2$, $Var(Y|\mathcal{H}_0)=\frac{2}{N}\xi^4$ and $Var(Y|\mathcal{H}_1)=\frac{2}{N}(\gamma+1)\xi^4$, respectively. Here, $\gamma$ is the received Signal-to-Noise Ratio (SNR) from the PT at the SG, and we have $\gamma=h^{\textrm{PT}}P_{\textrm{PT}}/\xi^2$ with the primary transmit power $P_{\textrm{PT}}$. Given the channel bandwidth $W$, sensing time $\tau^s$ and detection threshold $\epsilon$, we have $N=W\tau^s$ and according to~\cite{4489760, 6253200},
\begin{equation}
  \label{eq:detection}
  \left\{
 \begin{array}{ll}
 p^f(\tau^s, \epsilon) = Q\left(\displaystyle\frac{\epsilon-\xi^2}{\sqrt{2}\xi^2}\sqrt{W\tau^s}\right),\\
 p^d(\tau^s, \epsilon) = Q\left(\displaystyle\frac{\epsilon-(1+\gamma)\xi^2}{\sqrt{2}(1+\gamma)\xi^2}\sqrt{W\tau^s}\right),
 \end{array}\right.
\end{equation}
where $Q(x)=(1/\sqrt{2\pi})\int_{x}^{\infty}e^{-t^2/2}dt$ is known as the Q-function.\vspace*{-3.5mm}

\subsection{Transmission Based on Energy Harvesting and Backscattering}
The STs select their operation mode according to the channel detection result provided by the SG. When the channel is detected to be busy, an ST chooses to either harvest energy from the PT signals or backscatter for its own transmission. Otherwise, the ST can choose to transmit data using the energy harvested during the energy-harvesting phase. To address the conflict over channel usage among the STs, the TDMA mechanism is adopted by the CR network in its MAC layer, and the SG is responsible for synchronizing the phases of sensing, harvesting/backscattering and transmission among the STs. Since the performance of the STs depends on the accuracy of the channel sensing result, we also need to explicitly consider the impact of the sensing error on the ST operation in the three phases. Let $\tau^h_k$ denote the time length that is allocated to ST $k$ for energy harvesting when the channel is detected as busy. Then, the expected RF energy that is harvested from the PT by ST $k$ during the energy-harvesting phase $\tau^h_k$ is
\begin{equation}
 \label{eq:energy_harvested}
 E^H_k(\tau^s, \epsilon, \tau^h_k)=p_1p^d(\tau^s, \epsilon)\tau^h_k\delta h^{\textrm{PT}}_kP_{\textrm{PT}},
\end{equation}
where $p_1$ is the probability for the channel to be busy (i.e., hypothesis $\mathcal{H}_1$), $\delta$ is the energy harvesting efficiency ratio ($0\le\delta\le1$) and $h^{\textrm{PT}}_k$ is the channel power gain from the PT to ST $k$. Note that in (\ref{eq:energy_harvested}) we omit the case of hypothesis $\mathcal{H}_0$, since no energy can be sufficiently harvested when a false alarm happens and the channel is actually idle.

Let $\tau^b_k$ denote the length of time allocated to ST $k$ for backscattering. Since the backscattering bitrate is determined by the built-in backscatter circuit~\cite{Liu:2013:ABW:2486001.2486015}, we consider that the backscattering bitrate of ST $k$ is fixed as $\overline{r}^b_k$ during the backscattering phase $\tau^b_k$. Note that when a false alarm happens, ST $k$ cannot effectively backscatter due to the absence of the primary signals. Therefore, we can express the expected backscattering rate for ST $k$ during $\tau^b_k$ as follows:
\begin{equation}
 \label{eq:tx_backscatter}
 r^b_k(\tau^s, \epsilon)=p_1p^d(\tau^s, \epsilon)\overline{r}^b_k.
\end{equation}

Alternatively, when the channel is detected as idle and ST $k$ decides to perform active data transmission, its transmission rate in the opportunistic transmission phase depends on the available energy that is harvested during the energy harvesting phase. Since the circuit power consumption is not negligible for active transmission, we consider that each ST has to provide a fixed power $P^c$ for powering the circuit~\cite{7876867,7937935}.
Let $\tau^t_k$ denote the length of time allocated to ST $k$ for active data transmission. Consider that the STs adopt an opportunistic transmission policy by
sustaining the active transmission during $T$ with the expected level of harvested energy during the same length of period. From (\ref{eq:energy_harvested}), the expected power that ST $k$ achieves during $\tau^t_k$ is
\begin{equation}
 \label{eq:tx_power}
 P_k(\tau^s, \epsilon, \tau^h_k, \tau^t_k) = \frac{E^H_k(\tau^s, \epsilon, \tau^h_k)-P^c\tau_k^t}{\tau_k^t}=\frac{p_1p^d(\tau^s, \epsilon)\tau^h_k\delta h^{\textrm{PT}}_{k}P_{\textrm{PT}}}{\tau^t_k}-P^c,
\end{equation}
which is naturally accompanied by the supply-power constraint $P_k(\tau^s, \epsilon, \tau^h_k, \tau^t_k)\ge 0$.

Let $\sigma^2_k$ denote the AWGN power over the link from ST $k$ to the SG, and $h_k$ denote the corresponding channel power gain. Then, by taking into consideration the impact of false alarm and miss detection of the PT signals, the expected active transmission rate during $\tau^t_k$ is
\begin{equation}
 \label{eq:tx_rate_idle}
 \begin{array}{ll}
 r^t_k(\tau^s, \epsilon, \tau^h_k, \tau^t_k)\!=&\!p_0(1\!-\!p^f(\tau^s, \epsilon))\kappa_kW\log_2\left(1\!-\!P^c\!+\!\displaystyle\frac{h_kP_k(\tau^s, \epsilon, \tau^h_k, \tau^t_k)}{\sigma^2_k}\right)\\
 &+p_1(1\!-\!p^d(\tau^s, \epsilon))\kappa_kW\log_2\left(1\!-\!P^c\!+\!
 \displaystyle\frac{h_kP_k(\tau^s, \epsilon, \tau^h_k, \tau^t_k)}{h^{\textrm{PT}}P_{\textrm{PT}}\!+\!\sigma^2_k}\right),
 \end{array}
\end{equation}
where $p_0$ is the probability for the channel to be idle, $p_1$ is the probability for the channel to be busy, $\kappa_k$ is the transmission efficiency ratio ($0\le\kappa_k\le1$), $h^{\textrm{PT}}$ is the channel power gain from the PT to the SG, $h_k$ is the power gain for the secondary link $k$ and $P_k(\tau^s, \epsilon, \tau^h_k, \tau^t_k)$ is the expected transmit power given in (\ref{eq:tx_power}). On the right-hand-side of (\ref{eq:tx_rate_idle}), the first term represents the bitrate achieved by ST $k$ when the channel is correctly detected as idle, and the second term represents the bitrate achieved by ST $k$ when miss detection happens.

Let $T$ denote the total length of one time slot for the secondary network. After the length of the sensing phase $\tau^s$ is determined by the SG, the STs jointly determine the allocation of the sub-time slots, $\tau^h_k$, $\tau^b_k$ and $\tau^t_k$, within the accessible time range $[0, T-\tau^s]$ in the corresponding channel state. Following the TDMA mechanism, in either the transmitting mode or the backscattering mode, only one ST is allowed to operate over the channel at any time instance. Then, the sub-time slot allocation for ST $k$ ($1\!\le\!k\!\le\!K$) has to satisfy the feasibility constraints $\sum_{k=1}^{K}\tau^t_k\le(T-\tau^s)$, $\sum_{k=1}^{K}\tau^b_k\le(T-\tau^s)$ and $\tau^h_k+\tau^b_k\le(T-\tau^s)$. Let ${s}_k\!=\!(\tau^h_k, \tau^t_k, \tau^b_k)$ denote ST $k$'s individual choice for sub-time slot allocation. Then, given a pair of the sensing parameters $(\tau^s, \epsilon)$ set by the SG, the transmission time scheduling problem for ST $k$ ($1\!\le\!k\!\le\!K$) can be formulated as follows:

\noindent \textbf{ST $k$'s utility optimization problem} is to find a strategy vector $s^*_k=(\tau^{h,*}_k, \tau^{t,*}_k,\tau^{b,*}_k)$ such that
\begin{subequations}\label{eq:opt:local}
\begin{align}
\tag{\ref{eq:opt:local}}
 s^*_k=\arg\max\limits_{s_k} & \Big(u_k(s_k; \tau^s, \epsilon)=\tau^b_k r^b_k(\tau^s, \epsilon) + \tau^t_k r^t_k(\tau^h_k,\tau^t_k; \tau^s, \epsilon)\Big),\\
\label{eq_opt:a}
 \textrm{s.t. } & \sum\limits_{i=1}^{K}\tau^b_i\le(T-\tau^s), \sum\limits_{i=1}^{K}\tau^t_i\le(T-\tau^s),\\
\label{eq_opt:b}
 &\tau^b_k + \tau^h_k\le(T-\tau^s), \tau^h_k\ge0, \tau^t_k\ge0, \tau^b_k\ge0,\\
\label{eq_opt:c}
 & p_1p^d(\tau^s,\epsilon)\delta h_k^{\textrm{PT}}P_{{\textrm{PT}}}\tau_k^h-P^c\tau_k^t\ge0,
\end{align}
\end{subequations}
where $r^b_k(\tau^s, \epsilon)$ and $r^t_k(\tau^h_k,\tau^t_k; \tau^s, \epsilon)$ are given in (\ref{eq:tx_backscatter}) and (\ref{eq:tx_rate_idle}), respectively. (\ref{eq_opt:a}) defines a set of common constraints that are shared by all the STs. It is worth noting that the two inequalities in
(\ref{eq_opt:a}), namely, $\sum_{i=1}^{K}\tau^b_i\le(T-\tau^s)$ and $\sum_{i=1}^{K}\tau^t_i\le(T-\tau^s)$, represent the constraints at state \emph{Busy} and state \emph{Idle}, respectively.
Let $\mathbf{s}^{\textrm{ST}}\!=\![s_1,\ldots,s_K]^{\top}$ denote the joint strategy vector for time resource allocation, and $s^{\textrm{ST}}_{-k}$ denote the joint strategies chosen by the adversaries of ST $k$. Then, from (\ref{eq_opt:a}), we note that the local strategy searching space of ST $k$ is determined by the adversaries' strategy $s^{\textrm{ST}}_{-k}$. (\ref{eq_opt:b}) provides the feasibility constraints. (\ref{eq_opt:c}) is the linear supply-power constraint derived based on the discussion about (\ref{eq:tx_power}).

\section{Stackelberg Game for Time Resource Allocation}
\label{sec_stackelberg}
Based on the system model given in Section~\ref{sec_system_model}, now we are ready to introduce an interference pricing mechanism for the SG to control the time resource allocation process among the STs. In this section, we will first provide the Stackelberg game-based mathematical model of the interaction between the SG and the STs. Then, with the backward induction-based analysis, we will present a series of discoveries regarding the properties of the game.\vspace*{-3.5mm}

\subsection{Stackelberg Game Formulation}
From the perspective of the PT, a low interference level, hence a low miss detection probability $1\!-\!p^d(\tau^s,\epsilon)$ is expected for the secondary network. By (\ref{eq:detection}), the PT naturally prefers a long sensing phase and a small detection threshold. In contrast, given the constraint on the harvested energy, the STs prefer to extend their transmit phase as long as possible. Since with imperfect channel detection, the interference from the STs cannot be completely eliminated, we consider that the PT is able to tolerate a certain level of interference, provided that the STs pay compensation, i.e., price, for the interference that they cause in the harvest-then-transmit mode. Thereby, we consider that the SG works on behalf of the primary network and is able to collect the payments from the STs for the interference that they cause to the PT. For each ST, the interference is measured in the time fraction of colliding with the PT. To properly encourage or curb the primary channel usage by the STs, the gateway is allowed to adaptively choose the sensing time $\tau^s$, detection testing threshold $\epsilon$ and uniform interference price. Let $\alpha$ denote the unit price of the interference time, then, the SG's expected revenue optimization problem can be formulated as follows:

\noindent \textbf{The SG's revenue optimization problem} is to find a strategy vector $s^*_0\!=\!(\alpha^*, \tau^{s,*}, \epsilon^*)$ such that
\begin{subequations}\label{eq:time:revenue}
\begin{align}
\tag{\ref{eq:time:revenue}}
s^*_0=\arg\max\limits_{s_0=(\alpha, \tau^s, \epsilon)} & \left( \theta_0(s_0; \mathbf{s}^{\textrm{ST}}) = \alpha p_1\sum\limits_{k=1}^{K}(1-p^d(\tau^s, \epsilon))\tau^t_k.\right)\\
\label{eq:time_revenue_a}
 \textrm{s.t.}\quad & 1-p^d(\tau^s, \epsilon)\le\overline{p}^m, \\
\label{eq:time_revenue_b}
 &  T\ge\tau^s\ge 0,\alpha\ge0, \overline{\epsilon}\ge\epsilon\ge\underline{\epsilon},
\end{align}
\end{subequations}
where (\ref{eq:time_revenue_a}) sets the constraint on the probability of miss detection allowed by the PT.

Meanwhile, after accounting for the payment made to the SG for interference (cf., (\ref{eq:time:revenue})), the individual goal of ST $k$ now becomes maximizing the net payoff for its transmission. From (\ref{eq:time_revenue_a}), we obtain $p^d(\tau^s, \epsilon)\ge1-\overline{p}^m$ and thus are able to relax the constraint in (\ref{eq_opt:c}) by replacing $p^d(\tau^s,\epsilon)$ therein with $1-\overline{p}^m$. Let $\nu$ denote the STs' valuation per unit transmission rate. Then, based on the local optimization problem of ST $k$ defined in (\ref{eq:opt:local}), we formulate the following expected payoff optimization problem for the STs:

\noindent \textbf{ST $k$'s payoff optimization problem} is to find a strategy vector $s^*_k\!=\!(\tau^{h,*}_k, \tau^{t,*}_k,\tau^{b,*}_k)$ such that
\begin{subequations}\label{eq:o_modifiedpt:local}
\begin{align}
\tag{\ref{eq:o_modifiedpt:local}}
s^*_k=\arg\max\limits_{s_k=(\tau^h_k,\tau^t_k,\tau^b_k)}& \Big(\theta_k(s_k;s_0)=\nu\left(\tau^b_k r^b_k(\tau^s, \epsilon) + \tau^t_k r^t_k(\tau^h_k,\tau^t_k; \tau^s, \epsilon)\right)-\alpha p_1(1-p^d(\tau^s, \epsilon))\tau^t_k\Big),\\
\label{eq_opt_modifiedpt_a}
\textrm{s.t.\;\;\quad} &  \sum\limits_{i=1}^{K}\tau^b_i\le(T-\tau^s), \sum\limits_{i=1}^{K}\tau^t_i\le(T-\tau^s),\\
\label{eq_opt_modifiedpt_b}
& \tau^b_k + \tau^h_k\le(T-\tau^s), \tau^h_k\ge0, \tau^t_k\ge0, \tau^b_k\ge0,\\
\label{eq_opt_modifiedpt_c}
 & p_1(1-\overline{p}^m)\delta h_k^{\textrm{PT}}P_{{\textrm{PT}}}\tau_k^h-P^c\tau_k^t\ge0.
\end{align}
\end{subequations}

The time scheduling problem described by (\ref{eq:time:revenue}) and (\ref{eq:o_modifiedpt:local}) can be naturally interpreted as a two-level decision making process. In the first level, the SG declares its selected values of the interference price, the sensing duration and the detection threshold. Then, following the SG's strategy, the STs negotiate among themselves about the allocation of the harvesting/backscattering and transmission sub-time slots. With such an allocation scheme, the problem of distributed time resource allocation can be formulated as a single-leader-multi-follower Stackelberg game.
\begin{Definition}[Stackelberg Game]
\label{def:Stackelberg_game}
 The two-level time scheduling game $\mathcal{G}$ is defined by a 3-tuple: $\langle \mathcal{K}=\{0, 1, \ldots, K\}, \mathcal{S}\!=\!\times\mathcal{S}_{k\in{\mathcal{K}}},
 \{\theta_k\}_{k\in{\mathcal{K}}}\rangle$, where player $k\!=\!0$ is the single leader (i.e., the SG), whose strategy space is $\mathcal{S}_0\!=\!\{s_0=(\alpha, \tau^s,\epsilon) : \overline{\epsilon}\!\ge\!\epsilon\!\ge\!\underline{\epsilon},\alpha\!\ge\!0, T\!\ge\!\tau^s\!\ge\!0, 1\!-\!p^d(\tau^s, \epsilon)\!\le\!\overline{p}^m\}$, and player $k$ ($k\!=\!1,\ldots,K$) is one follower player (i.e., an ST), whose strategy space is $\mathcal{S}_k\!=\!\{s_k\!=\!(\tau^h_k, \tau^t_k, \tau^b_k): \tau^h_k\!\ge\!0, \tau^t_k\!\ge\!0, \tau^b_k\!\ge\!0, \tau^h_k\!+\!\tau^b_k\!\le\!(T\!-\!\tau^s), p_1(1\!-\!\overline{p}^m)\delta h_k^{\textrm{PT}}P_{{\textrm{PT}}}\tau_k^h\!-\!P^c\tau_k^t\!\ge\!0\}\!\cap\!\{s_k\!=\!(\tau^h_k, \tau^t_k, \tau^b_k):\sum_{j=1}^{K}\tau^t_j\!+\!\tau^s\le T, \sum_{j=1}^{K}\tau^b_j\!+\!\tau^s\!\le\!T\}$. Player $k$'s individual payoff $\theta_k$ is given by the objective function in (\ref{eq:time:revenue}) and (\ref{eq:o_modifiedpt:local}) for $k\!=\!0$ and $k\!\ne\!0$, respectively.
\end{Definition}

Based on Definition~\ref{def:Stackelberg_game}, we have the multi-follower sub-game among the STs in $\mathcal{G}$ as a 3-tuple: $\mathcal{G}^f\!=\!\langle \mathcal{K}^{\textrm{ST}}\!=\!\{1, \ldots,K\}, \mathcal{S}^{\textrm{ST}}\!=\!\times\mathcal{S}_k, \{\theta_k\}_{k=1}^{K}\rangle$. Then, we can define the Nash Equilibrium (NE) of $\mathcal{G}^f$ in the form of simultaneous best response as follows:
\begin{Definition}[Follower Sub-game NE]
\label{def:Nash_equilibrium}
Given the SG's strategy $s_0$, the parametric joint follower strategy $\mathbf{s}^{\textrm{ST}, *}(s_0)$ is an NE of $\mathcal{G}^f$ if $\forall k\!\in\!{\mathcal{K}^{\textrm{ST}}}$, the following condition holds $\forall s_k\!\in\!\mathcal{S}_k\left(s^{\textrm{ST}, *}_{-k}(s_0)\right)$:
 \begin{equation}
  \label{eq_equilibrium}
  \theta_k\left(s^*_k(s_0), s^{\textrm{ST}, *}_{-k}(s_0)\right)\ge\theta_k\left(s_k, s^{\textrm{ST}, *}_{-k}(s_0)\right).
 \end{equation}
\end{Definition}
Based on the follower sub-game NE given in Definition~\ref{def:Nash_equilibrium}, we can further define the SE of game $\mathcal{G}$ as the following sub-game perfect NE:
\begin{Definition}[SE]
\label{def:Stackelberg_equilibrium}
 $\mathbf{s}^*=({s^*_k})_{k=0}^{K}$ is the SE of game $\mathcal{G}$ if the following inequality is satisfied:
 \begin{equation}
  \label{eq_SE}
   \theta_0\left(s^*_0, \mathbf{s}^{\textrm{ST},*}(s^*_0)\right)\ge\theta_0\left(s_0, \mathbf{s}^{\textrm{ST},*}(s_0)\right),
 \end{equation}
 where $\forall s_0\in \mathcal{S}_0$, $\mathbf{s}^{\textrm{ST},*}(s_0)$ is one of the rational reactions of the followers satisfying (\ref{eq_equilibrium}).
\end{Definition}

From Definition~\ref{def:Nash_equilibrium}, we note that for any player $k\!\ne\!0$ in the follower sub-game $\mathcal{G}^f$, its strategy space $\mathcal{S}_k$ depends on the joint adversaries' strategy, $s^{\textrm{ST}}_{-k}=(s_i)_{i\in{\mathcal{K}^{\textrm{ST}}},i\ne k}$. Namely, $s_k\in\mathcal{S}_k(s_{-k})$ is a set-valued map which depends on the shared, rival-strategy dependent constraints given in (\ref{eq_opt_modifiedpt_a}). Therefore, the problem of NE seeking for game $\mathcal{G}^f$ becomes a Generalized NE (GNE) problem~\cite{Facchinei2010}. Furthermore, to obtain the joint SE strategy $\mathbf{s}^*$, the followers' rational reaction mapping, $\mathbf{s}^{f,*}(s_0)$, is required to be established for the follower sub-game given any leader strategy $s_0$.
Then, the problem of SE seeking in $\mathcal{G}$ becomes a bilevel programming problem with multiple lower-level local optimization problems and a single upper-level optimization problem~\cite{dempe2002foundations}. Thereby, we analyze the properties of the SE in game $\mathcal{G}$ through backward induction by first investigating the properties of the NE in the follower sub-game $\mathcal{G}^f$.\vspace*{-3.5mm}

\subsection{Analysis of the Follower Sub-game}\label{sub_sec_follower_analysis}
Assume that the leader's strategy is fixed as $s_0\!=\!(\alpha, \tau^s, \epsilon)$. For conciseness, from now on we omit $s_0$ in ST $k$'s strategy space $\mathcal{S}_k(s^{\textrm{ST}}_{-k}, s_0)$ and payoff function $\theta_k(s_k, s^{\textrm{ST}}_{-k}, s_0)$ in the analysis of the follower sub-game. Then, we have the following properties in regard to $\mathcal{G}^f$:
\begin{Theorem}
 \label{thm_convex_game}
 The following properties hold with respect to the objective and constraint functions in ST $k$'s payoff optimization problem defined by (\ref{eq:o_modifiedpt:local}):
 \begin{itemize}
 \item[\bf{P1}:] $\mathcal{S}_k$ is convex and compact $\forall k\in\mathcal{K}^{\textrm{ST}}$, and for any feasible $s^{\textrm{ST}}_{-k}$, $\mathcal{S}_k(s^{\textrm{ST}}_{-k})$ is nonempty.
 \item[\bf{P2}:] $\forall k\!\in\!\mathcal{K}^{\textrm{ST}}$, the objective function $\theta_k(s_k, s^{\textrm{ST}}_{-k})$ given by (\ref{eq:o_modifiedpt:local}) is a twice continuously differentiable ($C^2$) concave function with respect to $s_k$.
\end{itemize}
\end{Theorem}

\begin{proof}
 See Appendix~\ref{Proof_thm_convex_game}.
\end{proof}

Theorem~\ref{thm_convex_game} indicates that for each ST, the local optimization problem in (\ref{eq:o_modifiedpt:local}) is a concave programming problem. Theorem~\ref{thm_convex_game} paves the way of resorting to the mathematical tool of Quasi-Variational Inequalities (QVI)~\cite{Facchinei2010} for showing the existence of the GNE in the follower sub-game. Before proceeding, we first provide the definition of the QVI problem as follows:
\begin{Definition}[VI~\cite{5447064}]
  \label{def_VI}
 Given a closed and convex set $\mathcal{S}\in\mathbb{R}^n$ and a gradient-based mapping $F:\mathcal{S}\rightarrow\mathbb{R}^n$, the VI problem denoted as $\mathop{\textrm{VI}}
 (\mathcal{S},F)$, consists of finding a vector $\mathbf{s}^*\in\mathcal{S}$, called a solution of the VI, such that:
 \begin{equation}
  \label{eq_vi_definition}
  (\mathbf{y}-\mathbf{s}^*)^TF(\mathbf{s}^*)\ge 0, \forall\mathbf{y}\in\mathcal{S}.
 \end{equation}
If the defining set $\mathcal{S}$ depends on the variable $\mathbf{s}$, i.e., $\mathbf{s}\in\mathcal{S}( \mathbf{s})$, then, $\mathop{\textrm{VI}}(\mathcal{S},F)$ is a QVI problem.
\end{Definition}

From Definition~\ref{def:Stackelberg_game}, we define $F^f\!=\!(-\nabla_{{s}_k}\theta_k(\mathbf{s}^{\textrm{ST}}))_{k=1}^{K}$ and obtain a corresponding QVI problem $\mathop{\textrm{VI}}(\mathcal{S}^{\textrm{ST}},F^f)$, where $\mathcal{S}^{\textrm{ST}}$ is given by the definition of the follower sub-game $\mathcal{G}^f$. Then, we have the following property that guarantees the equivalence between the solution to the reformulated QVI problem $\mathop{\textrm{VI}}(\mathcal{S}^{\textrm{ST}},F^f)$ and the GNE of the original follower sub-game $\mathcal{G}^f$:
\begin{Lemma}
\label{lemma_equivalence}
A joint follower strategy $\mathbf{s}^{\textrm{ST},*}$ is a GNE of the follower sub-game $\mathcal{G}^f$ if and only if it is a solution of the QVI problem $\mathop{\textrm{VI}}(\mathcal{S}^{\textrm{ST}},F^f)$.
\end{Lemma}
\begin{proof}
 With {\bf{P1}} and {\bf{P2}} in Theorem~\ref{thm_convex_game}, Lemma~\ref{lemma_equivalence} immediately follows Theorem 3.3 in~\cite{Facchinei2010}.
\end{proof}

By Lemma~\ref{lemma_equivalence}, to show the existence of the NE of sub-game $\mathcal{G}^f$, it suffices to show that the solution set to the QVI problem $\mathop{\textrm{VI}}(\mathcal{S}^{\textrm{ST}},F^f)$ is non-empty. Through inspecting the convexity and compactness of the strategy set and the monotonicity property of $F^f$, we obtain Theorem~\ref{thm_existence_NE_subgame}.
\begin{Theorem}
 \label{thm_existence_NE_subgame}
 For any feasible $s_0$, the follower sub-game $\mathcal{G}^f$ admits at least one GNE. Furthermore, let the GNE be denoted by $\mathbf{s}^{\textrm{ST},*}\!=\![s^*_1,\ldots,s^*_K]^{\top}$, then for ST $k\!\in\!\mathcal{K}^{\textrm{ST}}$, $\tau^{h,*}_k\!=\!T\!-\!\tau^s\!-\!\tau^{b,*}_k$.
\end{Theorem}
\begin{proof}
See Appendix~\ref{app_unique_sub_game_GNE}.
\end{proof}

Theorem~\ref{thm_existence_NE_subgame} shows that the STs tend to fully utilize the time fraction for channel state \emph{Busy} to backscatter or harvest energy. Then, we can remove one of the inter-dependent strategy variables $\tau_k^h$ and $\tau_k^b$ and obtain $s_k=(\tau^h_k=\!T\!-\!\tau^s\!-\!\tau^{b}_k, \tau^t_k,\tau^b_k)$ without affecting the sub-game NE as the joint solution to (\ref{eq:o_modifiedpt:local}). Since for each ST, the local optimization problem in (\ref{eq:o_modifiedpt:local}) is a concave programming problem, we can derive the GNE of the follower sub-game through solving the concatenated Karush-Kuhn-Tucker (KKT) conditions of the local problems $\forall k\!\in\!\mathcal{K}^\textrm{ST}$. Let $G(\mathbf{s}^\textrm{ST})$ denote the vector of constraints that are jointly determined by $\mathbf{s}^\textrm{ST}$, and $Z_k(s_k)$ denote the vector of constraints that depend only on the local strategy $s_k$. Then, from (\ref{eq:o_modifiedpt:local}) we have
\begin{equation}
 \label{eq_joint_constraint}
 G(\mathbf{s}^\textrm{ST})=\begin{bmatrix}
    G^1(\mathbf{s}^\textrm{ST})\\
    G^2(\mathbf{s}^\textrm{ST})
   \end{bmatrix}
  =\begin{bmatrix}
    \sum\limits_{k=1}^{K}\tau_k^t-(T-\tau^s)\\
    \sum\limits_{k=1}^{K}\tau_k^b-(T-\tau^s)
   \end{bmatrix},
\end{equation}
and
\begin{equation}
 \label{eq_local_constraint}
 Z_k(s_k)=\begin{bmatrix}
    Z_k^1(s_k)\\
    Z_k^2(s_k)\\
    Z_k^3(s_k)
  \end{bmatrix}^{\top}
  =\begin{bmatrix}
    -\tau_k^t\\
    -\tau_k^b\\
    P^c\tau_k^t-p_1(1-\overline{p}^m)\delta h_k^{\textrm{PT}}P_{{\textrm{PT}}}(T-\tau^s-\tau_k^b)
   \end{bmatrix}^{\top}.
\end{equation}
Let $\pmb\lambda_k$ and $\pmb\mu_k$ denote the KKT multiplier vector for $G$ and $Z_k$ in the local optimization problem of ST $k$, respectively. Then, for ST $k$ the KKT conditions are as follows:
\begin{align}
 \label{eq_kkt_1}
 &\nabla_{s_k}\theta_k(s_k)-\left(\nabla_{s_k}G(\mathbf{s}^\textrm{ST})\right)^{\top}\pmb\lambda_k-\left(\nabla_{s_k}Z_k(s_k)\right)^{\top}\pmb\mu_k=0,\\
  \label{eq_kkt_2}
 &\mathbf{0}\le\pmb\lambda_k\perp -G(\mathbf{s}^\textrm{ST})\ge 0,\\
  \label{eq_kkt_3}
 &\mathbf{0}\le\pmb\mu_k\perp -Z_k({s}_k)\ge 0,
\end{align}
where (\ref{eq_kkt_2}) and (\ref{eq_kkt_3}) provide the complementary conditions and the operator $\perp$ represents component-wise orthogonality. Namely, for two vectors $\mathbf{x}$ and $\mathbf{y}$, $\mathbf{x}\perp\mathbf{y}\Leftrightarrow\mathbf{x}_i\mathbf{y}_i\!=\!0, \forall i$. Observing $G(\mathbf{s}^{\textrm{ST}})$ and
$Z_k({s}_k)$, we note that all the constraint functions are affine.
Thereby, we can immediately find a feasible strategy $\tilde{s}_k=\left(\tau_k^t=\frac{(T-\tau^s)}{2K}, \tau^b_k=\min\left(\frac{(T-\tau^s)}{2K},T-\tau^s-\frac{P^c\tau_k^t}{p_1(1-\overline{p}^m)\delta h^{\textrm{PT}}_kP_{\textrm{PT}}}\right)\right)$ that guarantees $\forall k\ge1$, $G^i\le0$ and $Z^j_{k}\le0$ for all the constraint indices $i$ and $j$ at $\tilde{s}_k$. Then, according to the Slater's theorem (cf. Chapter 5.2.3 of~\cite{boyd2004convex}), $\tilde{s}_k$ is in the relative interior of the strategy domain and satisfies the Slater's condition.
Therefore, strong duality holds for the Lagrangian of the local optimization problem in (\ref{eq:o_modifiedpt:local}) and the KKT conditions given by (\ref{eq_kkt_1})-(\ref{eq_kkt_3}) provide both the necessary and sufficient condition for an optimal solution to (\ref{eq:o_modifiedpt:local}). Then, we have Lemma~\ref{lemma_kkt_solution}.
\begin{Lemma}
 \label{lemma_kkt_solution}
 If $\left(\mathbf{s}^{\textrm{ST},*}, (\pmb\lambda^*_k)_{k=1}^K, (\pmb\mu^*_k)_{k=1}^K\right)$ solves the concatenated KKT system given by (\ref{eq_kkt_1})-(\ref{eq_kkt_3}), then $\mathbf{s}^{\textrm{ST},*}$ is a GNE point of the follower sub-game $\mathcal{G}^f$.
\end{Lemma}
\begin{proof}
 With {\bf{P1}} and {\bf{P2}} in Theorem~\ref{thm_convex_game} and the strong duality of the Lagrangian function corresponding to (\ref{eq_kkt_1}) shown above, Lemma~\ref{lemma_kkt_solution} immediately follows Theorem 4.6 of~\cite{Facchinei2010}.
\end{proof}

Lemma \ref{lemma_kkt_solution} naturally leads to the idea of deriving the follower sub-game equilibria through identifying the solution of the concatenated local KKT systems given by (\ref{eq_kkt_1})-(\ref{eq_kkt_3}) for all $k\in\mathcal{K}^{\textrm{ST}}$. Further inspection into the structure of $\mathcal{G}^f$ reveals that a simplified form of the solution to the concatenated KKT system can be obtained. This relies on showing that the follower sub-game $\mathcal{G}^f$ is an exact potential game~\cite{MONDERER1996124}:
\begin{Lemma}
 \label{lemma_potential}
 The follower game $\mathcal{G}^f$ is an exact potential game with the following potential function
 \begin{equation}
  \label{eq_potential_func}
  \phi(\mathbf{s}^{\textrm{ST}})=\sum_{k=1}^{K}\theta_k(s_k,s^{\textrm{ST}}_{-k}).
 \end{equation}
\end{Lemma}
\begin{proof}
 From (\ref{eq:o_modifiedpt:local}) we note that $\theta_k(\mathbf{s}^{\textrm{ST}})$ only depends on the local strategy $s_k$. Then, from (\ref{eq_potential_func}), $\forall s_k, s'_k\in\mathcal{S}^{\textrm{ST}}_k(s^{\textrm{ST}}_{-k})$ the following holds with any given $s^{\textrm{ST}}_{-k}\in\mathcal{S}^{\textrm{ST}}_{-k}$:
 \begin{equation}
  \label{eq_potential_equiv}
  \begin{array}{ll}
  \phi(s_k,s^{\textrm{ST}}_{-k})\!-\!\phi(s'_k, s^{\textrm{ST}}_{-k})\!=\!\left(\theta_k(s_k)\!+\!\sum_{j\ne k}\theta_j(s_j)\right)\!-\!\left(\theta_k(s'_k)\!+\!\sum_{j\ne k}\theta_j(s_j)\right)
  \!=\!\theta_k(s_k)-\theta_k(s'_k).
  \end{array}\nonumber
 \end{equation}
By the definition of the potential game~\cite{MONDERER1996124}, $\mathcal{G}^f$ is an exact potential game.
\end{proof}

Based on Lemma~\ref{lemma_potential}, we are able to convert the multi-player, non-cooperative sub-game $\mathcal{G}^f$ into a single optimization problem and obtain Lemma~\ref{lemma_socail_optimal}.
\begin{Lemma}
  \label{lemma_socail_optimal}
  The solution to the concatenated local KKT systems given by (\ref{eq_kkt_1})-(\ref{eq_kkt_3}), $\left({s}^*_k, \pmb\lambda^*_k, \pmb\mu^*_k\right)$, $\forall k\!\in\!\mathcal{K}^{\textrm{ST}}$, is also the socially optimal NE in $\mathcal{G}^f$. Furthermore, $\pmb\lambda^*_k\!=\!\pmb\lambda^*$, $\forall k\!\in\!\mathcal{K}^{\textrm{ST}}$.
\end{Lemma}
\begin{proof}
  See Appendix~\ref{app_socail_optimal}.
\end{proof}
Lemmas~\ref{lemma_kkt_solution}-\ref{lemma_socail_optimal} make it possible to introduce the Lagrangian-based analysis of the NE in $\mathcal{G}^f$. Based on Lemma~\ref{lemma_socail_optimal}, we can further verify the uniqueness of the NE in $\mathcal{G}^f$ and obtain Theorem~\ref{thm_social_optimal}:
\begin{Theorem}
 \label{thm_social_optimal}
The follower sub-game $\mathcal{G}^f$ admits a unique NE. Namely, the concatenated KKT system given by (\ref{eq_kkt_1})-(\ref{eq_kkt_3}) has a unique solution in the form of $(\mathbf{s}^{\textrm{ST},*},\pmb\lambda^*, \pmb\mu^*_1,\ldots, \pmb\mu^*_K)$.
\end{Theorem}
\begin{proof}
   See Appendix~\ref{app_socail_optimal}.
\end{proof}\vspace*{-3.5mm}

\subsection{Analysis of Stackelberg Equilibria in Game $\mathcal{G}$}
\label{sec_existence_ne}
By Theorem~\ref{thm_social_optimal}, $\mathcal{G}^f$ admits a unique GNE given any $s_0$. Let $\mathcal{E}(s_0)$ denote such a GNE mapping from $s_0$ and $\mathop{\textrm{gph}}{\mathcal{E}(s_0)}\!=\!\{(s_0,\mathbf{s}^{\textrm{ST}}):\mathbf{s}^{\textrm{ST}}\!=\!\mathcal{E}(s_0)\}$ denote the graph of $\mathcal{E}(s_0)$. Then, by Definition~\ref{def:Stackelberg_game}, the feasible region of the SE in $\mathcal{G}$ is $\Omega(s_0,\mathbf{s}^{\textrm{ST}})\!=\!\mathcal{S}_0\cap\mathop{\textrm{gph}}{\mathcal{E}(s_0)}$. After including the potential function-based KKT system given by (\ref{eq_kkt_num_1})-(\ref{eq_kkt_num_3}) into the leader's optimization problem in (\ref{eq:time:revenue}), the SE in game $\mathcal{G}$ is equivalent to the global solution of the following Mathematical Programming with Equilibrium Constraints (MPEC) problem~\cite{dempe2002foundations}:
\begin{subequations}\label{eq_time_revenue_reformulated}
\begin{align}
\tag{\ref{eq_time_revenue_reformulated}}
s^*_0=\arg\max\limits_{s_0=(\alpha, \tau^s, \epsilon)} & \left( \theta_0(s_0, \mathbf{s}^{\textrm{ST}}) = \alpha p_1\sum\limits_{k=1}^{K}(1-p^d(\tau^s, \epsilon))\tau^t_k.\right)\\
\label{eq_time_revenue_reformulated_1}
 \textrm{s.t.}\quad & 1-p^d(\tau^s, \epsilon)\le\overline{p}^m, \\
\label{eq_time_revenue_reformulated_2}
 &  T\ge\tau^s\ge 0,\alpha\ge0, \overline{\epsilon}\ge\epsilon\ge\underline{\epsilon},\\
\label{eq_kkt_num_reform_1}
 & \mathbf{s}^{\textrm{ST}}=\mathcal{E}(s_0),
\end{align}
\end{subequations}
where (\ref{eq_time_revenue_reformulated_1})-(\ref{eq_time_revenue_reformulated_2}) defines $\mathcal{S}_0$, and $\mathcal{E}(s_0)$ in (\ref{eq_kkt_num_reform_1}) is the parametric solution to the KKT system given by (\ref{eq_kkt_num_1})-(\ref{eq_kkt_num_3}). Since the objective function in (\ref{eq_time_revenue_reformulated}) is continuous in $s_0$ and coercive in $\alpha$, namely, $\theta_0(s_0, \mathbf{s}^{\textrm{ST}})\!\rightarrow\!\infty$ if $\alpha\!\rightarrow\!\infty$, by the well-known Weierstrass Theorem~\cite{sundaram1996first}, at least one global optimal solution in (\ref{eq_time_revenue_reformulated}) exists if $\Omega(s_0,\mathbf{s}^{\textrm{ST}})$ is non-empty and closed, and the objective function $\theta_0(s_0, \mathbf{s}^{\textrm{ST}}(s_0))$ is continuous in $s_0$. Therefore, we are able to develop the following theorem (cf. Theorems 5.1 in~\cite{dempe2002foundations}) regarding the SE in game $\mathcal{G}$.
\begin{Theorem}
 \label{thm_existence_GNE}
 Game $\mathcal{G}$ admits at least one global SE as defined by (\ref{eq_SE}).
\end{Theorem}
\begin{proof}
See Appendix~\ref{app_proof_SE_existence}.
\end{proof}

By replacing the implicit function $\mathbf{s}^{\textrm{ST}}\!=\!\mathcal{E}(s_0)$ in (\ref{eq_kkt_num_reform_1}) with the KKT system given in (\ref{eq_kkt_num_1})-(\ref{eq_kkt_num_3}), (\ref{eq_time_revenue_reformulated}) reduces the bilevel programming problem for SE searching into a single-level problem. However, we note from (\ref{eq:o_modifiedpt:local}) that $\theta_k(s_k,s_0)$ is a transcendental function of $\tau^t_k$. Then, a closed-form solution to the KKT system in (\ref{eq_kkt_num_1})-(\ref{eq_kkt_num_3}) does not exists. Moreover, due to the complementary conditions in (\ref{eq_kkt_num_2}) and (\ref{eq_kkt_num_3}), standard qualification conditions are violated everywhere in (\ref{eq_time_revenue_reformulated}) (see also Theorem 5.11 in~\cite{dempe2002foundations}). Therefore, (\ref{eq_time_revenue_reformulated}) is a non-convex problem to which the classical KKT-based analysis does not apply. Fortunately, from the proof of Theorem~\ref{thm_existence_GNE}, we know that the followers' parametric NE $\mathbf{s}^{\textrm{ST}}\!=\!\mathcal{E}(s_0)$ is piecewise continuously differentiable ($PC^1$), hence directionally differentiable (cf. Corollary 4.1 in~\cite{dempe2002foundations}). Thereby, instead of relying on heuristic method for SE searching (cf. \cite{7942105}), in what follows, we are able to implement a directional ascent-based method for the SE computation, which allows the follower sub-game NE to be solved as a nested problem in a distributed manner.\vspace*{-3.5mm}

\section{Distributed Approach for Computing Stackelberg Equilibrium}
\label{sec_SE_searching_method}
\subsection{Directional Ascent Method for SE Searching}
\label{sub_directional_SE_search}
\begin{algorithm}[t]
  \begin{small}
  \caption{Directional ascent method for SE searching}
  \begin{algorithmic}[1]
    \REQUIRE
    Select a feasible $s_0(t=0)$, choose updating coefficient $\rho\in(0,1)$.
    \WHILE {the condition $\Vert s_0(t\!+\!1)\!-\!s_0(t)\Vert\le\chi_0$ is not satisfied for a given precision $\chi_0\!>\!0$}
      \STATE Compute a direction vector $\mathbf{r}(t)$, $\Vert \mathbf{r}(t)\Vert\!\le\!1$ such that, $\exists d(t)<0, d(t)\in\mathbb{R}$
      \begin{equation}
        \label{eq_update_condition}
        \theta'_0(s_0(t),\mathcal{E}(s_0(t)); \mathbf{r}(t))\ge-d(t), \; \nabla_{s_0}Z_0(s_0(t))\le-Z_0(s_0(t))+d(t),\nonumber
      \end{equation}
      where $\theta'_0(s_0(t),\mathcal{E}(s_0(t)); \mathbf{r}(t))$ is the directional derivative with respect to $\mathbf{r}(t)$. For vector $\mathbf{r}$ at strategy $s_0$, we have
      \begin{equation}
        \label{eq_direction_derivative}
        \theta'_0(s_0,\mathcal{E}(s_0); \mathbf{r})\!=\!\nabla_{s_0}\theta_0(s_0,\mathcal{E}(s_0))\mathbf{r}+\nabla_{\mathbf{s}^{\textrm{ST}}}\theta_0(s_0,\mathcal{E}(s_0))\mathcal{E}'(s_0; \mathbf{r}).
      \end{equation}
      \STATE Choose a step size $\beta(t)$ such that $s_0(t+1)=s_0(t)+\beta(t)\mathbf{r}(t)$ and
      \begin{equation}
        \label{eq_update_leader}
        \theta_0(s_0(t+1),\mathcal{E}(s_0(t+1))\ge \theta_0(s_0(t),\mathcal{E}(s_0(t)))-\rho\beta(t)d(t),\;  Z_0(s_0(t+1))\le0.\nonumber
      \end{equation}
      \STATE Set $t\leftarrow t+1$ and compute $\mathbf{s}^{\textrm{ST}}(t)=\mathcal{E}(s_0(t))$.
    \ENDWHILE
  \end{algorithmic}
  \label{alg_SE}
\end{small}
\end{algorithm}

Now, with the directional differentiability of the implicit function $\mathbf{s}^{\textrm{ST}}\!=\!\mathcal{E}(s_0)$, we apply the directional ascent algorithm (i.e. the prototypical algorithm proposed in~\cite{Dempe1996}) to solve the MPEC problem defined by (\ref{eq_time_revenue_reformulated}). Let $Z_0(s_0)\!=\!1-p^d(\tau^s, \epsilon)\!-\!\overline{p}^m$ denote the constraint function given in (\ref{eq:time_revenue_a}). Then, the directional ascent algorithm can be described in Algorithm~\ref{alg_SE}. Here, we note that prototypical method given by Algorithm~\ref{alg_SE} in itself does not designate a way of either finding the direction vector $\mathbf{r}(t)$ or finding the sub-game NE $\mathcal{E}(s_0(t))$ at $s_0(t)$. For the convenience of discussion, we temporarily assume that the value of $\mathcal{E}(s_0)$ and its corresponding set of Lagrange multipliers $(\pmb\lambda, \pmb\mu)$ in the solution to (\ref{eq_kkt_num_1})-(\ref{eq_kkt_num_3}) are accessible for every $s_0$. Let $\tilde{G}(\mathbf{s}^{\textrm{ST}})$ be the vector of all the lower-level constraints given by (\ref{eq_opt_modifiedpt_a})-(\ref{eq_opt_modifiedpt_c}), which is formed through concatenating $G(\mathbf{s}^{\textrm{ST}})$ in (\ref{eq_joint_constraint}) and $Z_k({s}_k)$, $\forall k\!\in\!\mathcal{K}^{\textrm{ST}}$ in (\ref{eq_local_constraint}). Let $I_0(s_0, \mathbf{s}^{\textrm{ST}})\!=\!\{i: \tilde{G}_i(s_0, \mathbf{s}^{\textrm{ST}})\!=\!0\}$ be the set of active lower-level constraints (see also Definition~\ref{def_CRCQ}). Then, according to Theorem 3.4 in~\cite{Dempe1996} (cf. Theorem 5.4 in~\cite{dempe2002foundations}), finding the directional vector $\mathbf{r}(t)$, the intermediate scalar parameter $d(t)$ and the directional derivative of $\theta'_0(s_0,\mathcal{E}(s_0); \mathbf{r}(t))$ in Algorithm~\ref{alg_SE} is equivalent to solving the following linear programming problem with $\mathbf{s}^{\textrm{ST}}=\mathcal{E}(s_0)$ and $I_0$:
\begin{subequations}\label{eq_directional_gradient}
\begin{align}
\tag{\ref{eq_directional_gradient}}
(d^*, \mathbf{r}^*, \pmb\nu^*, \pmb\zeta_i^*)=& \arg\min\limits_{(d, \mathbf{r}, \pmb\nu, \pmb\zeta)} d\\
 \label{eq_constraint_lagragian_a}
 \textrm{s.t.}\quad & -\nabla^{\top}_{\mathbf{s}^{\textrm{ST}}}\theta_0(s_0,\mathbf{s}^{\textrm{ST}})\pmb\nu-\nabla^{\top}_{s_0}\theta_0(s_0,\mathbf{s}^{\textrm{ST}})\mathbf{r}\le d\\
 \label{eq_constraint_lagragian_b}
 &  \nabla^{\top}_{{s}_0}Z_0(s_0,\mathbf{s}^{\textrm{ST}})\mathbf{r}\le -Z_0(s_0,\mathbf{s}^{\textrm{ST}})+ d,\\
 \label{eq_constraint_lagragian}
 & -\nabla^2_{(\mathbf{s}^{\textrm{ST}})^2}L(s_0,\mathbf{s}^{\textrm{ST}},\pmb\lambda,\pmb\mu)\pmb\nu
 -\nabla^2_{\mathbf{s}^{\textrm{ST}}s_0}L(s_0,\mathbf{s}^{\textrm{ST}},\pmb\lambda,\pmb\mu)\mathbf{r}+
 \nabla_{\mathbf{s}^{\textrm{ST}}}\tilde{G}(\mathbf{s}^{\textrm{ST}})\pmb\zeta=\mathbf{0},\\
 &\nabla^{\top}_{\mathbf{s}^{\textrm{ST}}}\tilde{G}_i(\mathbf{s}^{\textrm{ST}})\pmb\nu={0}, \quad \forall i\in I_0,\\
 &\nabla^{\top}_{\mathbf{s}^{\textrm{ST}}}\tilde{G}_j(\mathbf{s}^{\textrm{ST}})\pmb\nu\le -\tilde{G}_j(\mathbf{s}^{\textrm{ST}})+d, \quad \forall j\notin I_0,\\
 &\zeta_i\ge0, i\in I_0, \quad \zeta_i=0, i\notin I_0, \quad  \Vert \mathbf{r}\Vert\le1.
\end{align}
\end{subequations}
For conciseness, we omit the iteration index $t$ in (\ref{eq_directional_gradient}). In (\ref{eq_constraint_lagragian}), $L(s_0,\mathbf{s}^{\textrm{ST}},\pmb\lambda, \pmb\mu)$ is the Lagrangian function for the lower-level problem as defined in (\ref{eq_Lagragian}), and $\pmb\zeta$ is a $(2\!+\!3K)$-dimentional vector. From (\ref{eq_joint_constraint}), (\ref{eq_local_constraint}) and (\ref{eq_Lagragian}), we note that $\nabla^2_{(\mathbf{s}^{\textrm{ST}})^2}G(\mathbf{s}^{\textrm{ST}})\!=\!\mathbf{0}$, $\nabla^2_{\mathbf{s}^{\textrm{ST}}s_0}G(\mathbf{s}^{\textrm{ST}})\!=\!\mathbf{0}$, $\nabla^2_{(\mathbf{s}^{\textrm{ST}})^2}Z_k(\mathbf{s}^{\textrm{ST}})\!=\!\mathbf{0}$ and $\nabla^2_{\mathbf{s}^{\textrm{ST}}s_0}Z_k(\mathbf{s}^{\textrm{ST}})\!=\!\mathbf{0}$, $\forall k\!\in\!\mathcal{K}^{\textrm{ST}}$. Then, we obtain $\nabla^2_{(\mathbf{s}^{\textrm{ST}})^2}L(s_0,\mathbf{s}^{\textrm{ST}},\pmb\lambda, \pmb\mu)\!=\!\nabla^2_{(\mathbf{s}^{\textrm{ST}})^2}\phi(s_0,\mathbf{s}^{\textrm{ST}})$ and $\nabla^2_{\mathbf{s}^{\textrm{ST}}s_0}L(s_0,\mathbf{s}^{\textrm{ST}},\pmb\lambda, \pmb\mu)\!=\!\nabla^2_{\mathbf{s}^{\textrm{ST}}s_0}\phi(s_0,\mathbf{s}^{\textrm{ST}})$ in (\ref{eq_constraint_lagragian}). As a result, the solution to (\ref{eq_directional_gradient}) does not require discovering the Lagrange multipliers $(\pmb\lambda, \pmb\mu)$ for a pair of strategies $(s_0,\mathbf{s}^{\textrm{ST}})$ in advance. Therefore, the problem given in (\ref{eq_directional_gradient}) can be effectively solved as long as the lower-level payoffs are available to the SG for strategy pair $(s_0,\mathbf{s}^{\textrm{ST}})$.

Following the proof of Theorem~\ref{thm_social_optimal}, we know that $\nabla_{s_k}\tilde{G}_i(\mathbf{s}^{\textrm{ST}})$ is constant. Also, given $(s_0,\mathbf{s}^{\textrm{ST}})$, as long as $\exists i, j\!\in\!\mathcal{K}^{\textrm{ST}}$ such that $Z_i^1(\mathbf{s}^{\textrm{ST}})\!\ne\!0$, $Z_j^2(\mathbf{s}^{\textrm{ST}})\!\ne\!0$ and $Z_i^3(\mathbf{s}^{\textrm{ST}})\!\ne\!0$, the gradients in the set $\{\nabla_{\mathbf{s}^{\textrm{ST}}}\tilde{G}_i(\mathbf{s}^{\textrm{ST}}): i\!\in\!I_0(s_0, \mathbf{s}^{\textrm{ST}})\}$ are linearly independent. When a feasible solution to (\ref{eq_directional_gradient}), $(d^*, \mathbf{r}^*, \pmb\nu^*, \pmb\zeta_i^*)$, is found with $d^*<0$, by Theorem 3.4 in~\cite{Dempe1996}, $\pmb\nu^*$ will be the directional derivative of the implicit function $\mathcal{E}'(s_0;\mathbf{r})$. Meanwhile, we can construct the following matrix:
\begin{equation}
  \label{eq_FRR}
  M=
  \begin{bmatrix}
    \nabla^2_{(\mathbf{s}^{\textrm{ST}})^2}\phi(s_0,\mathbf{s}^{\textrm{ST}}) & \nabla_{\mathbf{s}^{\textrm{ST}}}\tilde{G}_{i\in I_0(s_0, \mathbf{s}^{\textrm{ST}})}
    (\mathbf{s}^{\textrm{ST}}) & \nabla^2_{s_0\mathbf{s}^{\textrm{ST}}}\phi(s_0,\mathbf{s}^{\textrm{ST}})\\
    \nabla_{\mathbf{s}^{\textrm{ST}}}\tilde{G}_{i\in I_0(s_0, \mathbf{s}^{\textrm{ST}})}(\mathbf{s}^{\textrm{ST}}) & \mathbf{0} & \mathbf{0}
  \end{bmatrix}.
\end{equation}
It is tedious but easy to check that $\nabla^2_{s_0\mathbf{s}^{\textrm{ST}}}\phi(s_0,\mathbf{s}^{\textrm{ST}})$ is of full row rank. Then, following our discussion on the linear independency of the row vectors in $\nabla_{\mathbf{s}^{\textrm{ST}}}\tilde{G}_{i\in I_0(s_0, \mathbf{s}^{\textrm{ST}})}(\mathbf{s}^{\textrm{ST}})$, $M$ is of full row rank. By Theorem 6.1 in~\cite{dempe2002foundations}, since the conditions of SSOC, CRCQ (see Definitons~\ref{def_SSOC} and~\ref{def_CRCQ}) and full row rank of $M$ are satisfied, Algorithm~\ref{alg_SE} is guaranteed to converge to a local optimum solution to (\ref{eq_time_revenue_reformulated}). Therefore, the convergence to the local SE is guaranteed for Algorithm~\ref{alg_SE}.\vspace*{-3.5mm}

\subsection{Distributed Method for NE Searching in the Follower Sub-game}
\label{sub_sec_alg}
Algorithm~\ref{alg_SE} relies on the computation of the lower-level rational reactions $\mathbf{s}^{\textrm{ST}}\!=\!\mathcal{E}(s_0)$ at $s_0$ to determine the ascent direction. For the purpose of distributively finding the GNE of the follower sub-game $\mathcal{G}^f$, we introduce the regularized best-response algorithm (also known as proximal-response map) from~\cite{Facchinei2010} in Algorithm~\ref{alg_heuristic}. Algorithm~\ref{alg_heuristic} is a Gauss-Seidel-style algorithm based on a regularized objective function of the sub-problem in (\ref{eq:o_modifiedpt:local}) for iterative GNE searching. The convergence property of Algorithm~\ref{alg_heuristic} is proved in Theorem~\ref{thm_follower_convergence}.
\begin{algorithm}[t]
  \begin{small}
 \caption{Asynchronous proximal-response method for finding sub-game GNE}
 \begin{algorithmic}[1]
 \REQUIRE
 Select an initial strategy $\mathbf{s}^{\textrm{ST}}(t=0)=(s_1(0),\ldots, s_K(0))\in\mathcal{S}^{\textrm{ST}}$.
 \WHILE {the condition $\Vert \mathbf{s}^{\textrm{ST}}_0(t+1)-\mathbf{s}^{\textrm{ST}}_0(t)\Vert\le\chi^{\textrm{ST}}$ is not satisfied for a given $\chi^{\textrm{ST}}>0$}
  \FORALL {$k=1,\ldots,K$}
  \STATE Set the adversary joint strategies as
  \begin{equation}
   \label{eq_set_adversary}
    s^{\textrm{ST}}_{-k}(t)=(s_1(t+1),\ldots, s_{k-1}(t+1),s_{k+1}(t), \ldots, s_{K}(t)).
  \end{equation}

  \STATE Given $s^{\textrm{ST}}_{-k}(t)$, compute a local optimal solution $s_{k}(t+1)$:
  \begin{equation}
  \label{eq_best_response_update}
  \begin{array}{ll}
 s_{k}(t+1)=\arg\max\limits_{s_k}& \theta_k(s_k, s^{\textrm{ST}}_{-k}(t))-\displaystyle\frac{1}{2}\Vert s_k-s_k(t)\Vert^2,\\
 \qquad\qquad\qquad\quad\textrm{s.t.} & G(s_k, s^{\textrm{ST}}_{-k}(t))\le 0, \quad Z_k(s_k)\le 0.
 \end{array}
  \end{equation}
  \ENDFOR
  \STATE Set $t\leftarrow t+1$.
 \ENDWHILE
 \end{algorithmic}
 \label{alg_heuristic}
 \end{small}
\end{algorithm}

\begin{Theorem}[Convergence]
 \label{thm_follower_convergence}
 Algorithm \ref{alg_heuristic} converges to a GNE from any feasible $\mathbf{s}^{\textrm{ST}}(t=0)$.
\end{Theorem}
\begin{proof}
 The proof consists of two parts. In the first part, we employ the potential-game property of $\mathcal{G}^f$ and prove  by contradiction that if Algorithm~\ref{alg_heuristic} converges, it converges to a GNE of $\mathcal{G}^f$. In the second part, we exploit the monotonicity of $\mathop{\textrm{VI}}(\mathcal{S}^{\textrm{ST}},F)$ and show that Algorithm~\ref{alg_heuristic} is a contractive mapping and therefore always converges. See Appendix~\ref{app_proof_uniqueness} for the details.
\end{proof}
In Corollary~\ref{cor_jacobian_convergence}, we can further show that the proximal response also converges when the STs adopt a synchronous local strategy updating scheme.
\begin{Corollary}
 \label{cor_jacobian_convergence}
 The synchronous updating mechanism given by Algorithm \ref{alg_jacobian} (i.e., Jacobian best-response updating) converges to a GNE from any initial strategies $\mathbf{s}^{\textrm{ST}}(t=0)$.
\end{Corollary}
\begin{proof}
 See Appendix~\ref{app_proof_uniqueness}.
\end{proof}

\begin{Remark}
\rm
 The convergence of Algorithms~\ref{alg_heuristic} and~\ref{alg_jacobian} relies on the special structure of the utility function $\theta_k(s_k, s^{\textrm{ST}}_{-k})$, $\forall k\!\in\!\mathcal{K}^{\textrm{ST}}$. Namely, $\theta_k$ depends only on $s_k$ thus $\nabla_{\mathbf{s}^{\textrm{ST}}} F(\mathbf{s}^{\textrm{ST}})$ is a block diagonal matrix (see Appendix~\ref{app_unique_sub_game_GNE}). For a general case, it requires that $F(\mathbf{s}^{\textrm{ST}})$ in (\ref{eq_jacobian}) is a P(P$_0$)-property mapping~\cite{Facchinei2003}. Otherwise, the convergence conditions of Algorithms~\ref{alg_heuristic} and~\ref{alg_jacobian} are typically not known, and Algorithms~\ref{alg_heuristic} and~\ref{alg_jacobian} can be considered at most good heuristic~\cite{Facchinei2010}.
\end{Remark}\vspace*{-3.5mm}

\begin{algorithm}[t]
    \begin{small}
 \caption{Simultaneous best-response updating for finding GNE}
 \begin{algorithmic}[1]
 \REQUIRE
 Select an initial strategy $\mathbf{s}^{\textrm{ST}}(t=0)=(s_1(0),\ldots, s_K(0))\in\mathcal{S}^{\textrm{ST}}$.
 \WHILE {the termination criterion is not satisfied}
  \FORALL {$k=1,\ldots,K$}
  \STATE Given $\mathbf{s}^{\textrm{ST}}(t)$, compute a local optimal solution $s'_{k}$ according to (\ref{eq_best_response_update}).
  \ENDFOR
  \STATE $t\leftarrow t+1$, $s_{k}(t+1)=s'_{k}$.
 \ENDWHILE
 \end{algorithmic}
 \label{alg_jacobian}
\end{small}
\end{algorithm}

It is well known that a generic bilevel programming problem is strongly NP-hard, even when checking the local optimality of a strategy with both levels' objective functions being quadratic~\cite{7942105}. As a result, it is difficult to precisely determine the computational complexity of the proposed SE searching scheme described by Algorithms~\ref{alg_SE}-\ref{alg_jacobian}. Fortunately, we note that the upper-level directional ascent algorithm described by Algorithm~\ref{alg_SE} is executed at the SG. Since the SG  possesses sufficient computational power for performing the directional vector search in (\ref{eq_directional_gradient}) as well as the iterative gradient ascent, we only need to focus on the complexity of the distributed NE searching schemes for the follower subgame. For conciseness, we analyze the time complexity of Algorithm~\ref{alg_jacobian}, from which the similar result can be derived for Algorithm~\ref{alg_heuristic}. Following (\ref{eq_proof_1_14}) in Appendix~\ref{subsec_proof_jacobian_convergence}, let $\mathbf{s}(\cdot)$ denote the joint best response obtained from solving (\ref{eq_best_response_update}). Then, for any two feasible joint strategies of the STs, $\mathbf{y}$ and $\mathbf{z}$, there exists a constant $c\!\in\!(0,1)$ such that
\begin{equation}
  \label{eq_convergence_rate}
  \Vert \mathbf{s}(\mathbf{y})\!-\!\mathbf{s}(\mathbf{z})\Vert\!\le\!c\left\Vert \mathbf{y}-\mathbf{z}\right\Vert,
\end{equation}
where $\left\Vert\mathop{\textrm{Diag}}\left({\left\Vert-\nabla^2_{s^2_k}\theta_k(\tilde{\mathbf{x}})+I\right\Vert}^{-1}\right)_{k=1}^K\right\Vert\!\le\!c\!<\!1$, and according to the discussion about (\ref{eq_proof_1_11}), we have $\forall k\!\in\!\mathcal{K}, \exists \eta_k\!\in\!(0,1)$ such that $\tilde{x}_k=\eta_ks_k(\mathbf{z}) + (1-\eta_k)s_k(\mathbf{y})$.

With an initial strategy $\mathbf{s}^{\textrm{ST}}(0)$, we set the termination criterion in Algorithm~\ref{alg_jacobian} as $\Vert \mathbf{s}(\mathbf{s}^{\textrm{ST}}(t))- \mathbf{s}^{\textrm{ST},*}\Vert/\Vert \mathbf{s}^{\textrm{ST}}(0)- \mathbf{s}^{\textrm{ST},*}\Vert\le\chi^{\textrm{ST}}$, where $\chi^{\textrm{ST}}>0$ is the relative accuracy and $\mathbf{s}^{\textrm{ST},*}$ represents the follower subgame NE. Given a proper estimation of $c$ and an accuracy level $\chi^{\textrm{ST}}$ for algorithm termination, we can iteratively apply (\ref{eq_convergence_rate}) to the solution of (\ref{eq_best_response_update}) in Algorithm~\ref{alg_jacobian}. Then, we obtain the lower bound on the number of iterations $t$ in order for Algorithm~\ref{alg_jacobian} to converge as follows:
\begin{equation}
  \label{eq_time_complexity}
  t\ge\frac{\ln(1/\chi^{\textrm{ST}})}{\vert\ln c\vert}.
\end{equation}
\vspace*{-3.5mm}
\section{Simulation Results}
\label{sec_simulation}
\begin{table}[!t]
  \centering
  \caption{Main Parameters Used in the Simulation}\vspace*{-1mm}
  \scriptsize
 \begin{tabular}{|c c | c c | c c | c c | c c | c c|}
 \hline
 Parameter & Value & Parameter & Value & Parameter & Value & Parameter & Value & Parameter & Value & Parameter & Value\\
 \hline
 $p_0$ & $0.6$ & $W$ & $1$MHz & $T$ & 1s & AWGN  power & $-40$dBm & $\delta$  & $0.6$ & $\nu$ & $1$\\
 \hline
 $p_1$ & $0.4$ & $\gamma$ & $3$dB & $P_{\textrm{PT}}$ & 10W & $\xi^2$ & $0.01$ & $\kappa_k$ & $0.6$ & $P^c$ & -35dBm\\
 \hline
\end{tabular}
\label{table_parameter}\vspace*{-3.5mm}
\end{table}

\begin{figure*}[t]
\centering     
\subfigure[]{\label{fig_optimal_exposition_a}\includegraphics[width=.30\linewidth]{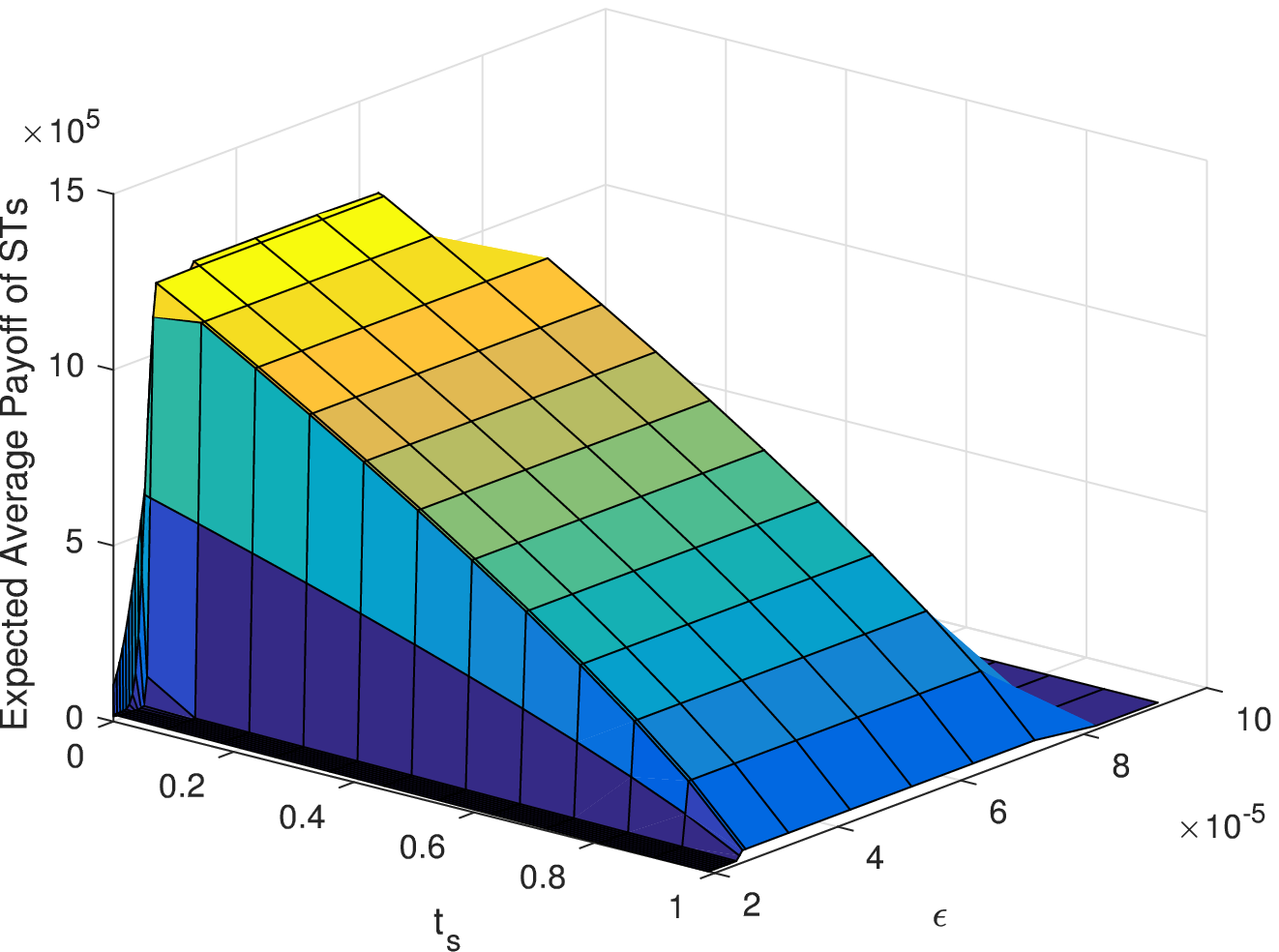}}
\subfigure[]{\label{fig_optimal_exposition_b}\includegraphics[width=.30\linewidth]{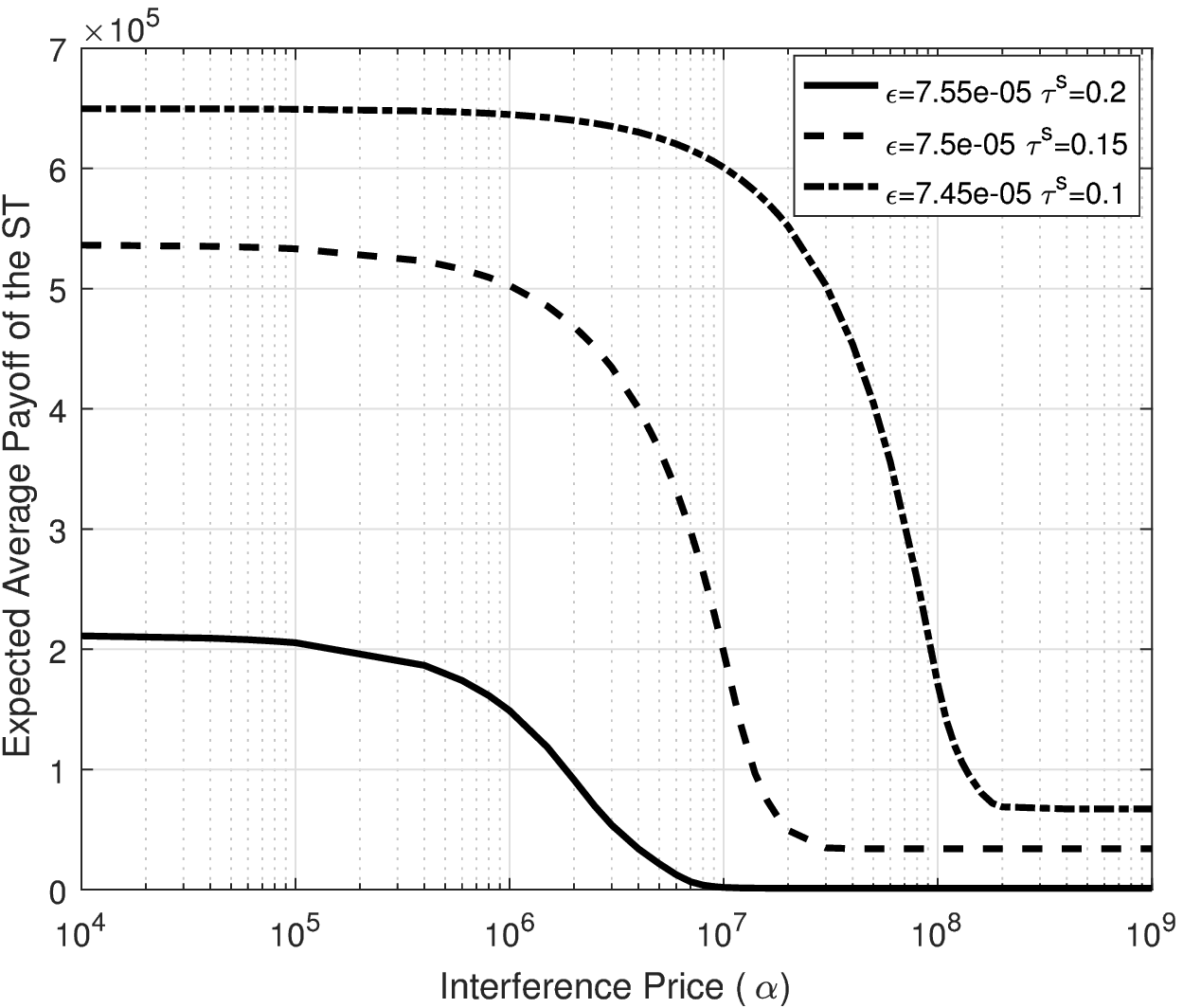}}
\subfigure[]{\label{fig_optimal_exposition_c}\includegraphics[width=.30\linewidth]{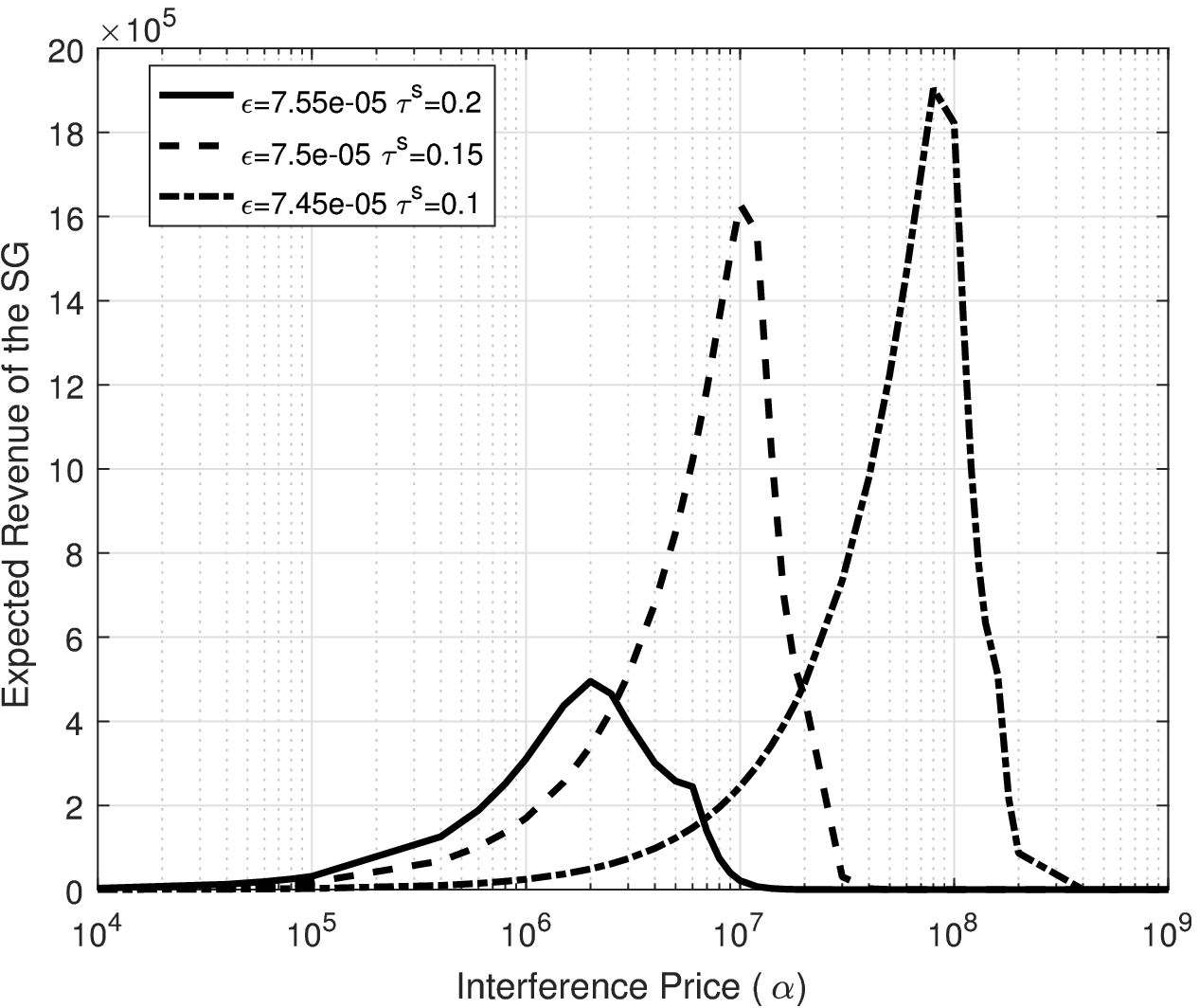}}\vspace*{-1mm}
\caption{(a) Expected average throughput of the STs with respect to the varied sensing time and detection threshold when the interference price is fixed as $\alpha\!=\!10^{8}$. (b) Expected average payoff of the STs vs. the varied interference price with different sensing strategies. (c) Expected revenue of the SG vs. the varied interference price with different sensing strategies.}
\label{fig_optimal_exposition}\vspace*{-3mm}
\end{figure*}
For the ease of exposition, we assume that the backscattering rates for the STs are the same, and the STs are randomly placed near the SG within a distance of $D\!\le\!30$m. We adopt a lognormal shadowing path loss model for the channel gains as $h_k=D_k^{-l}$, where $l$ is the path loss factor, $l=3.5$.  We employ the Monte Carlo simulations to approximate the node performance of the nodes, and the major parameters used in the simulation are listed in Table~\ref{table_parameter}. In our simulations, we first consider a secondary network with $5$ STs. Since the ST's throughput and payoff are a function of the SG's strategy $s_0=(\alpha, \tau^s,\epsilon)$, in Figure~\ref{fig_optimal_exposition}, we provide the graphical insight into the impact of the SG's sensing strategy and pricing strategy on the performance of the STs, respectively. From Figure~\ref{fig_optimal_exposition_a}, we observe that the STs' performance is more sensitive to the detection threshold $\epsilon$, since a small $\epsilon$ will result in the probability of false alarm $p^f$ sharply rising to 1, while a large $\epsilon$ will result in the probability of detection $p^d$ quickly falling to 0. On the other hand, as we expect, Figure~\ref{fig_optimal_exposition_b} shows that by adjusting the interference price, the SG can efficiently control the STs' usage of the idle time fraction in a time slot for direct transmission. An extremely high price will drive all the STs to completely evacuate from using the idle state for their transmission and operate only in the backscattering mode. Furthermore, by comparing Figures~\ref{fig_optimal_exposition_b} and \ref{fig_optimal_exposition_c}, we note that the maximum revenue of the SG will be reached at the loss of the ST's payoff, but before any ST refuses using any of the idle sub-time slot.

\begin{figure*}[t]
\centering     
\subfigure[]{\label{fig_performance_comparison_a}\includegraphics[width=.32\linewidth]{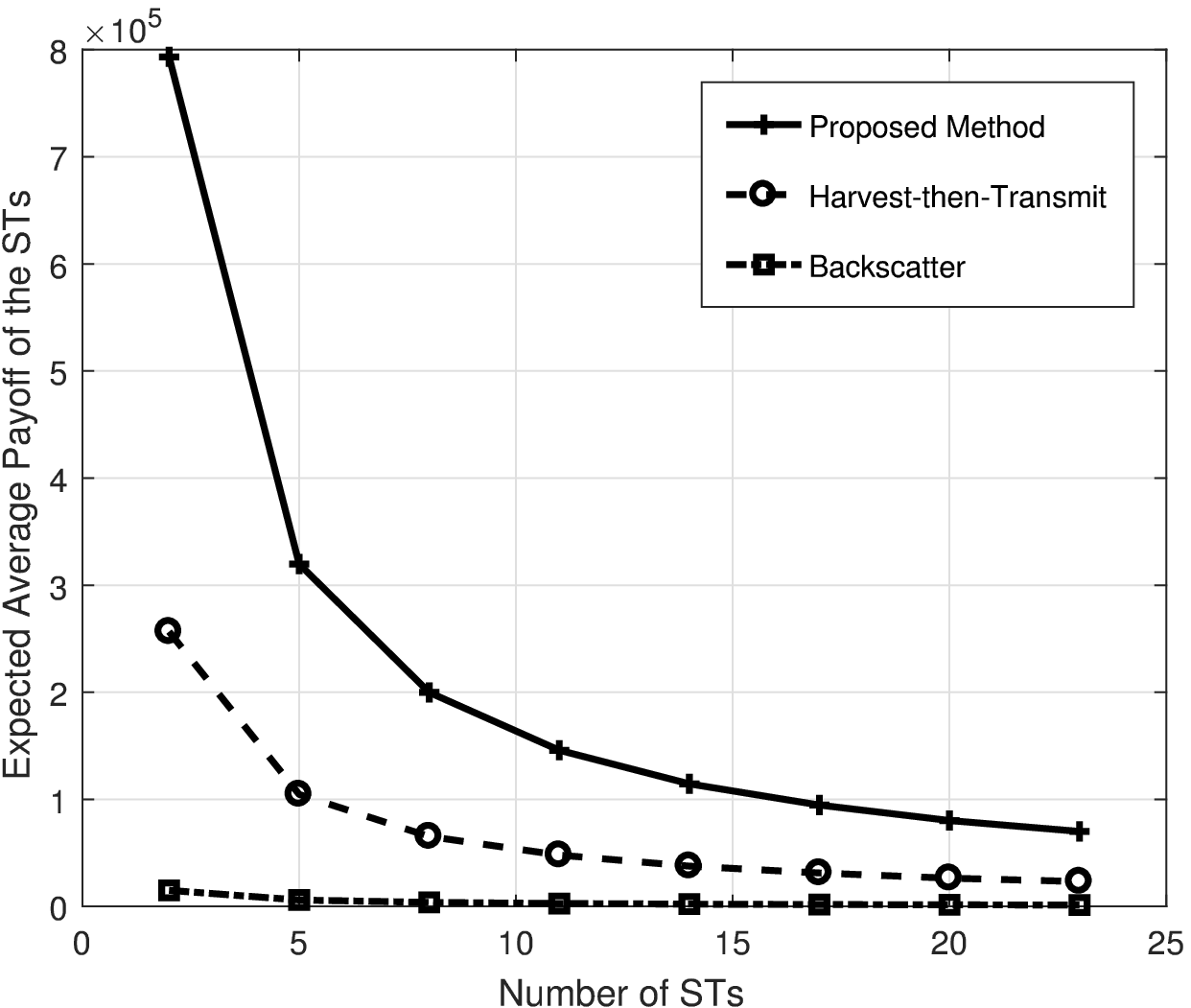}}\hspace{5mm}
\subfigure[]{\label{fig_performance_comparison_b}\includegraphics[width=.32\linewidth]{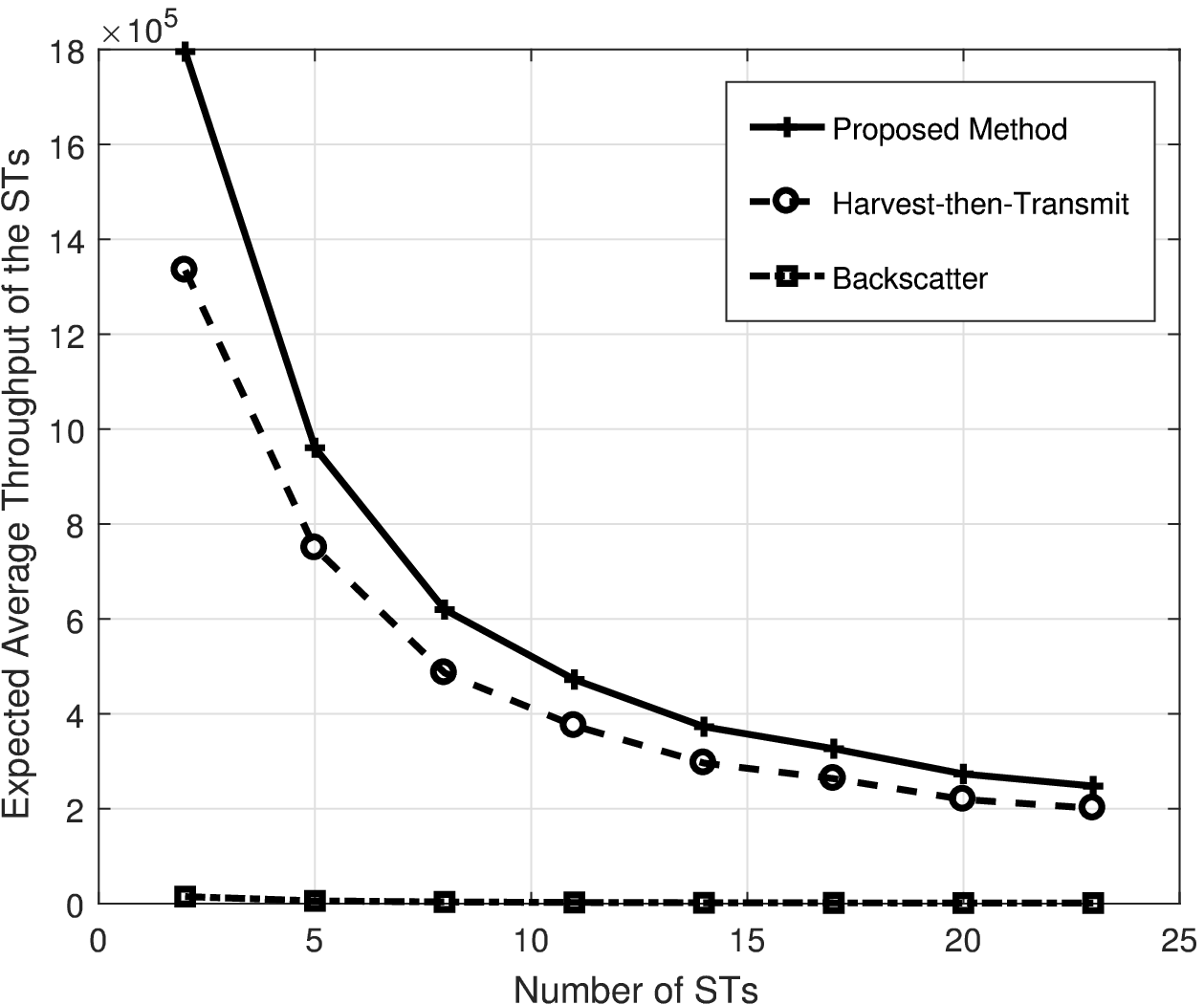}}
\subfigure[]{\label{fig_performance_comparison_c}\includegraphics[width=.32\linewidth]{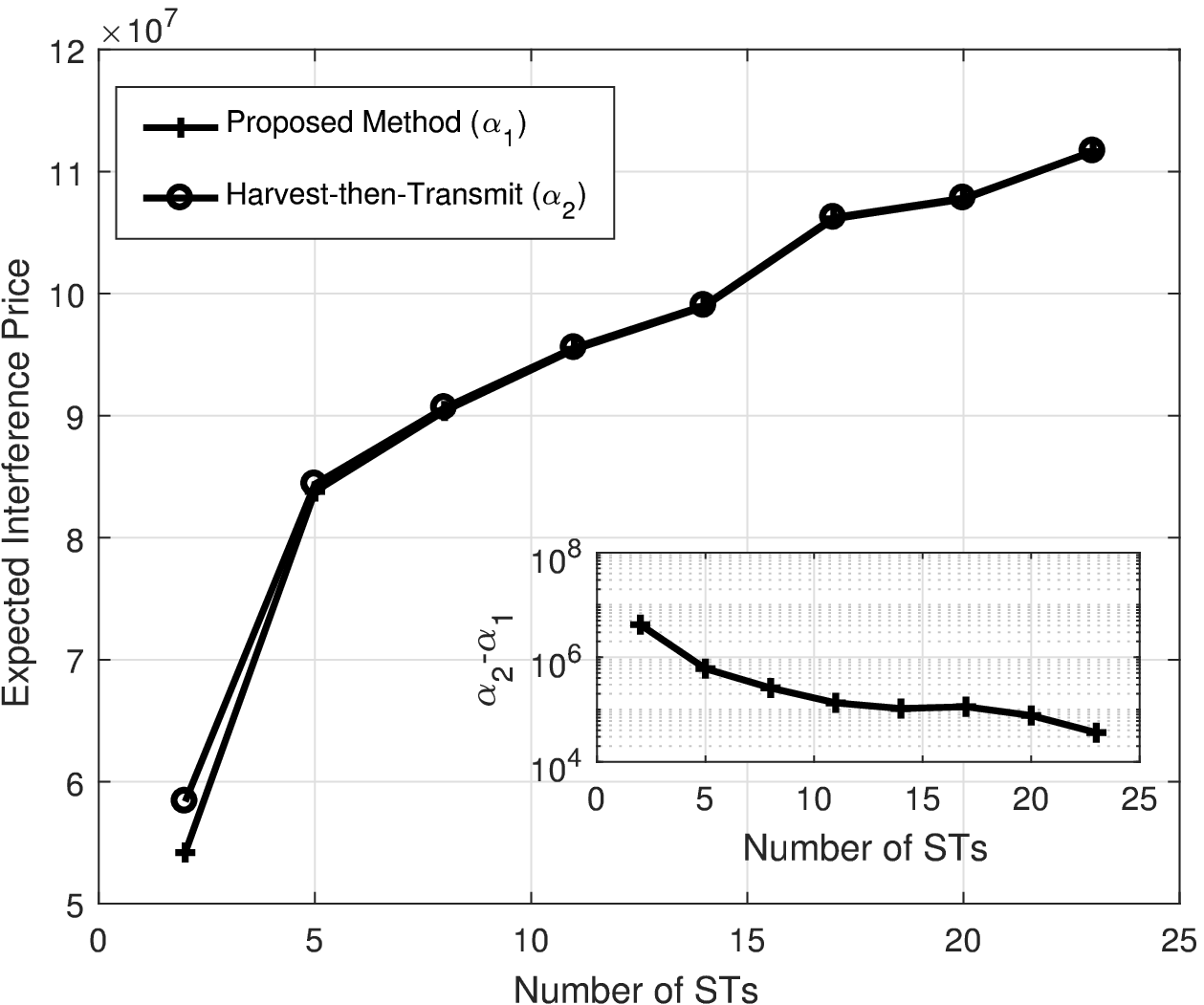}}\hspace{5mm}
\subfigure[]{\label{fig_performance_comparison_d}\includegraphics[width=.32\linewidth]{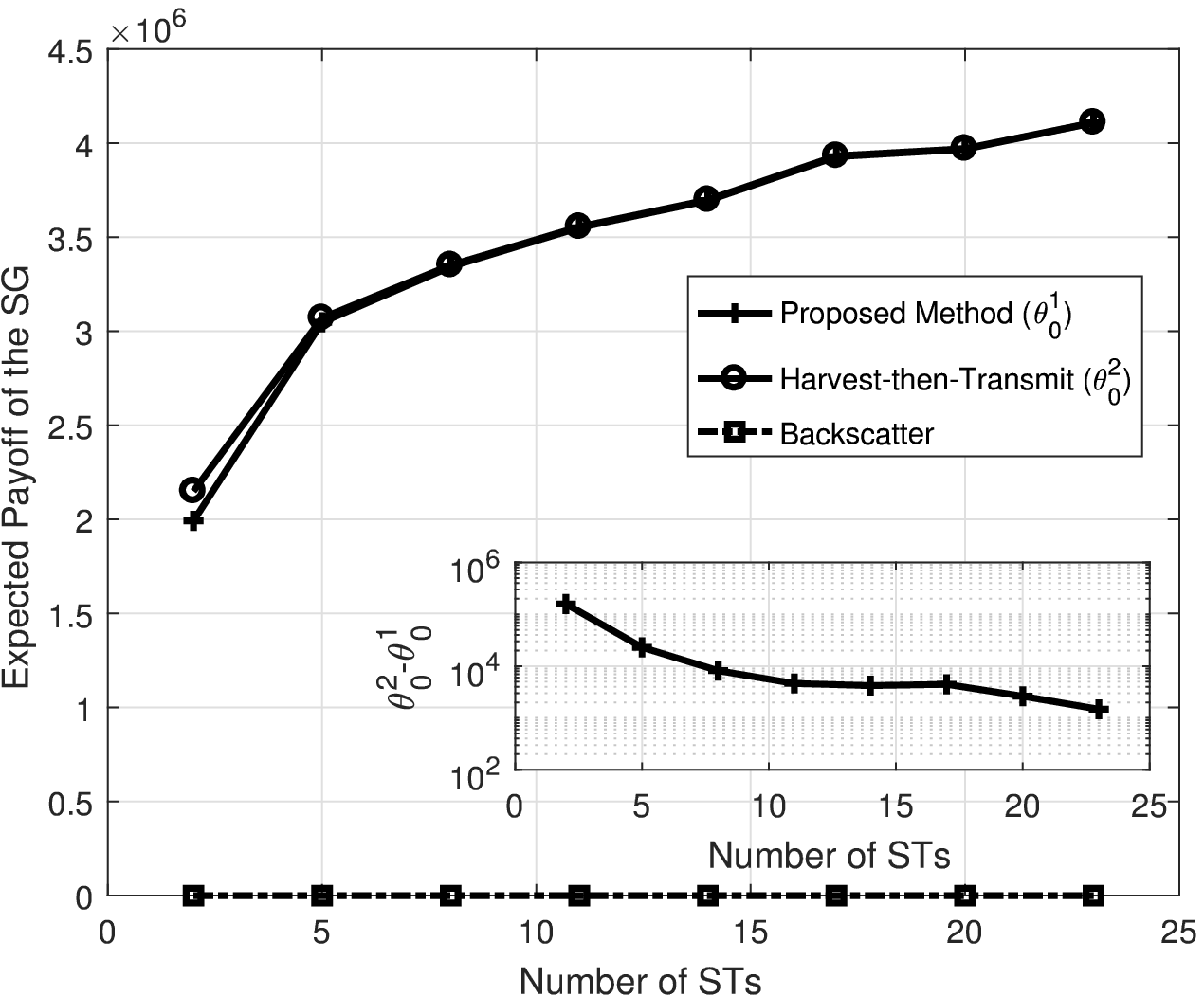}}\vspace*{-1mm}
\caption{Performance comparison for the proposed method, the harvest-then-transmit-only scheme and the backscatter-only scheme.
(a) Expected average payoff of the STs at the SE vs. the number of the STs. (b) Expected average throughput of the STs at the SE vs. the number of the STs. (c) Expected interference price set by the SG at the SE vs. the number of the STs. (d) Expected payoff of the SG at the SE vs. the number of the STs.}
\label{fig_performance_comparison}\vspace*{-3mm}
\end{figure*}

In Figure~\ref{fig_performance_comparison}, we compare the performance of the proposed algorithm with the performance of harvest-then-transmit-only and backscatter-only schemes in difference network scales. From Figures~\ref{fig_performance_comparison_a} and~\ref{fig_performance_comparison_b} we observe that the difference between the average payoff/throughput achieved by the proposed method and the harvest-then-transmit-only scheme is larger than the average throughput achieved by the backscatter-only scheme. This indicates that by adopting the hybrid transmit scheme, the STs have more advantage in negotiating the price with the SGs than with the harvest-then-transmit-only scheme. This phenomenon can also be observed in Figure~\ref{fig_performance_comparison_c}, since the equilibrium price asked by the SG in the harvest-then-transmit-only scheme is always slightly higher than that with the proposed method, although the performance of the former is significantly lower than that of the latter. This indicates that by adopting the hybrid transmission scheme, the STs' performance gain is larger than the sum of the performance of both the harvest-then-transmit-only and backscatter-only schemes.
Theoretically, with the proposed scheme, the STs are able to switch to the backscattering mode whenever the interference price exceeds the critical level. For the STs, in this situation operating in harvest-then-transmit mode will incur more payment due to interference. Therefore, completely staying in the backscattering mode will provide a better payoff. As a result, if the interference price is too high, the STs are always able to ``threaten'' to completely abstain from active transmission such that the SG receives zero payment. In return, this will discourage the SG from continuously increasing the interference price. By contrast, with the harvest-then-transmit-only scheme, the STs have no choice but continue their transmission when the interference price keeps rising, until some of the STs are forced out of play (i.e., stop transmitting) due to negative payoffs.

\begin{figure*}[t]
\centering     
\subfigure[]{\label{fig_probability_comparison_a}\includegraphics[width=.34\linewidth]{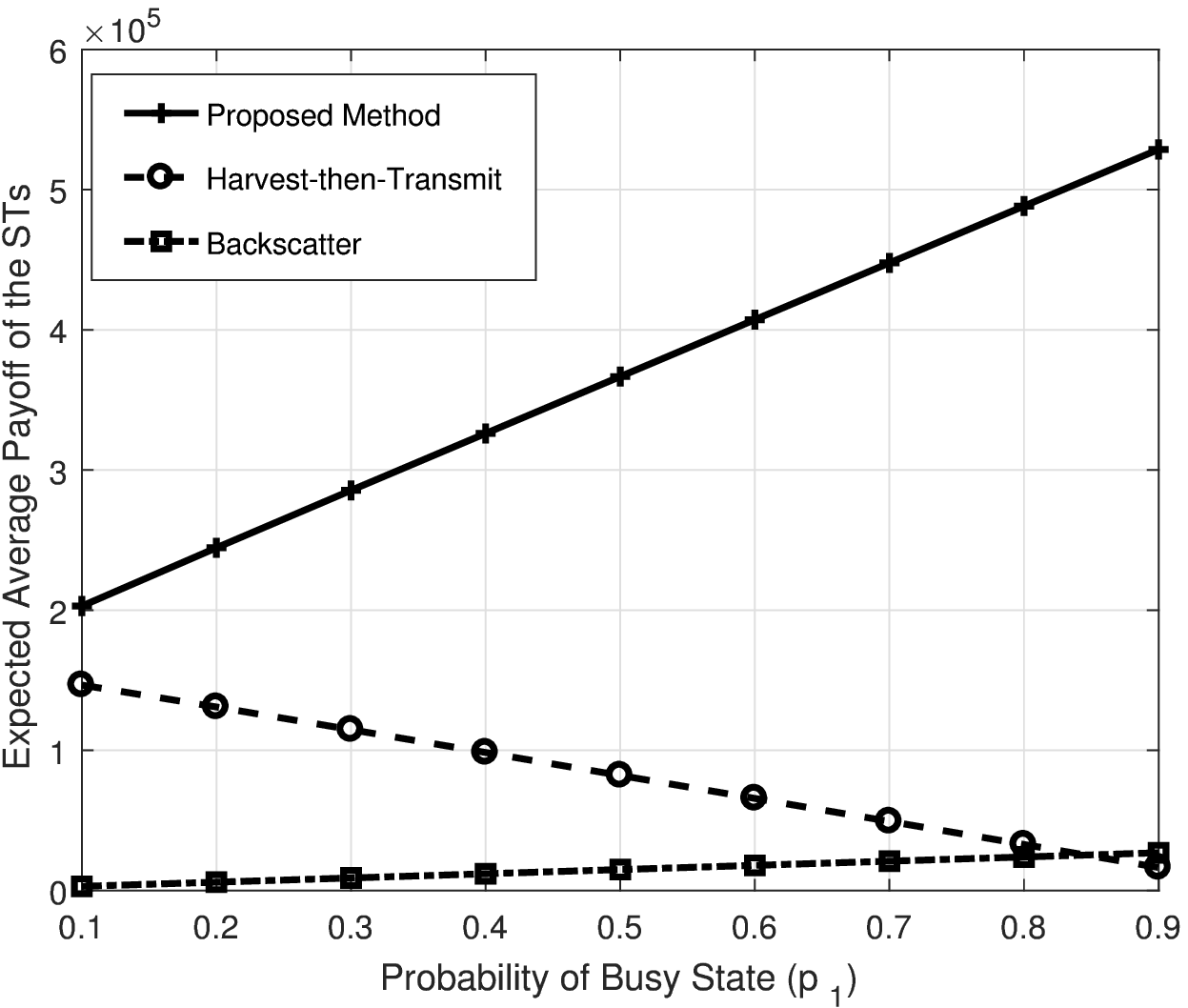}}\hspace{5mm}
\subfigure[]{\label{fig_probability_comparison_b}\includegraphics[width=.34\linewidth]{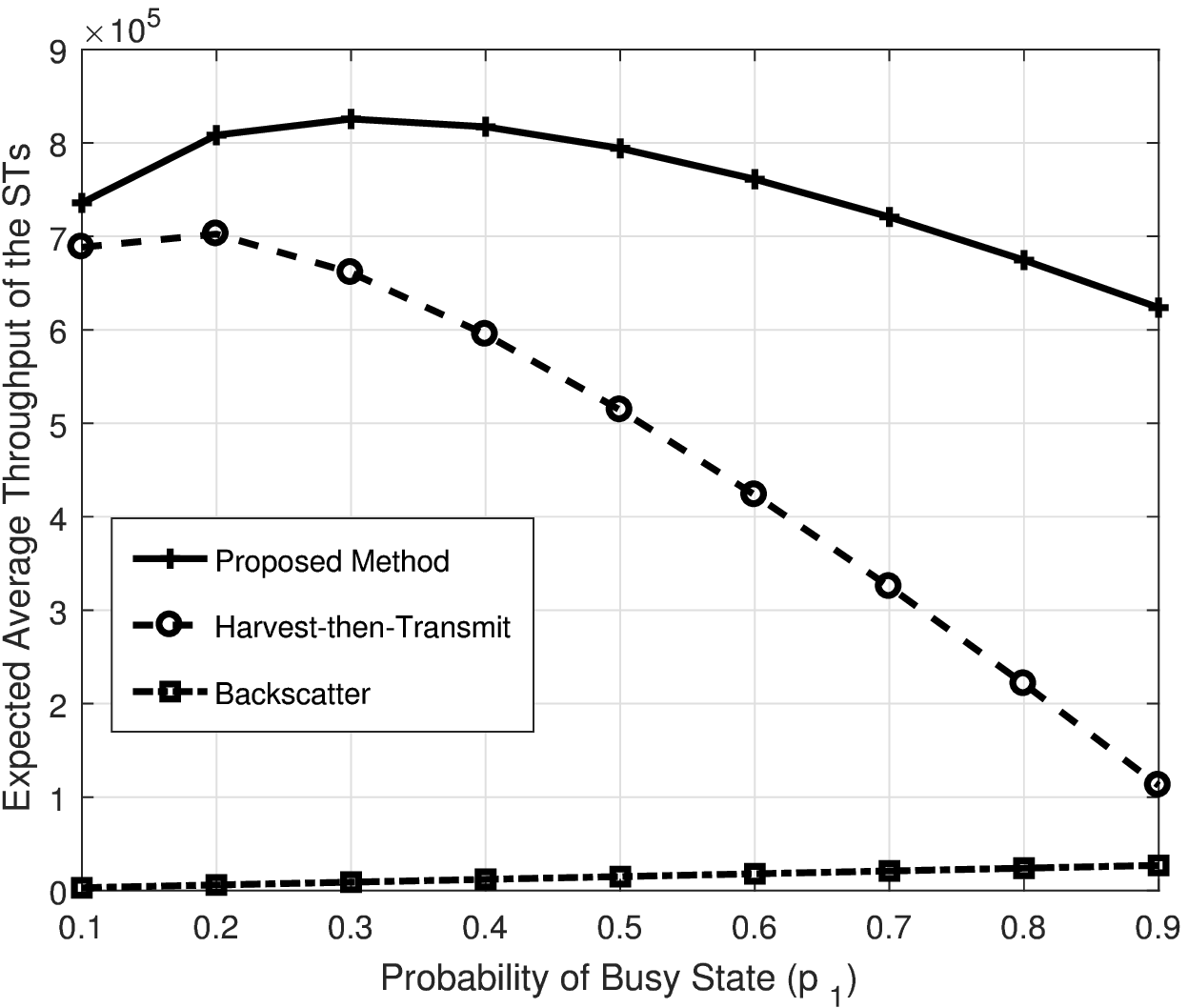}}
\subfigure[]{\label{fig_probability_comparison_c}\includegraphics[width=.34\linewidth]{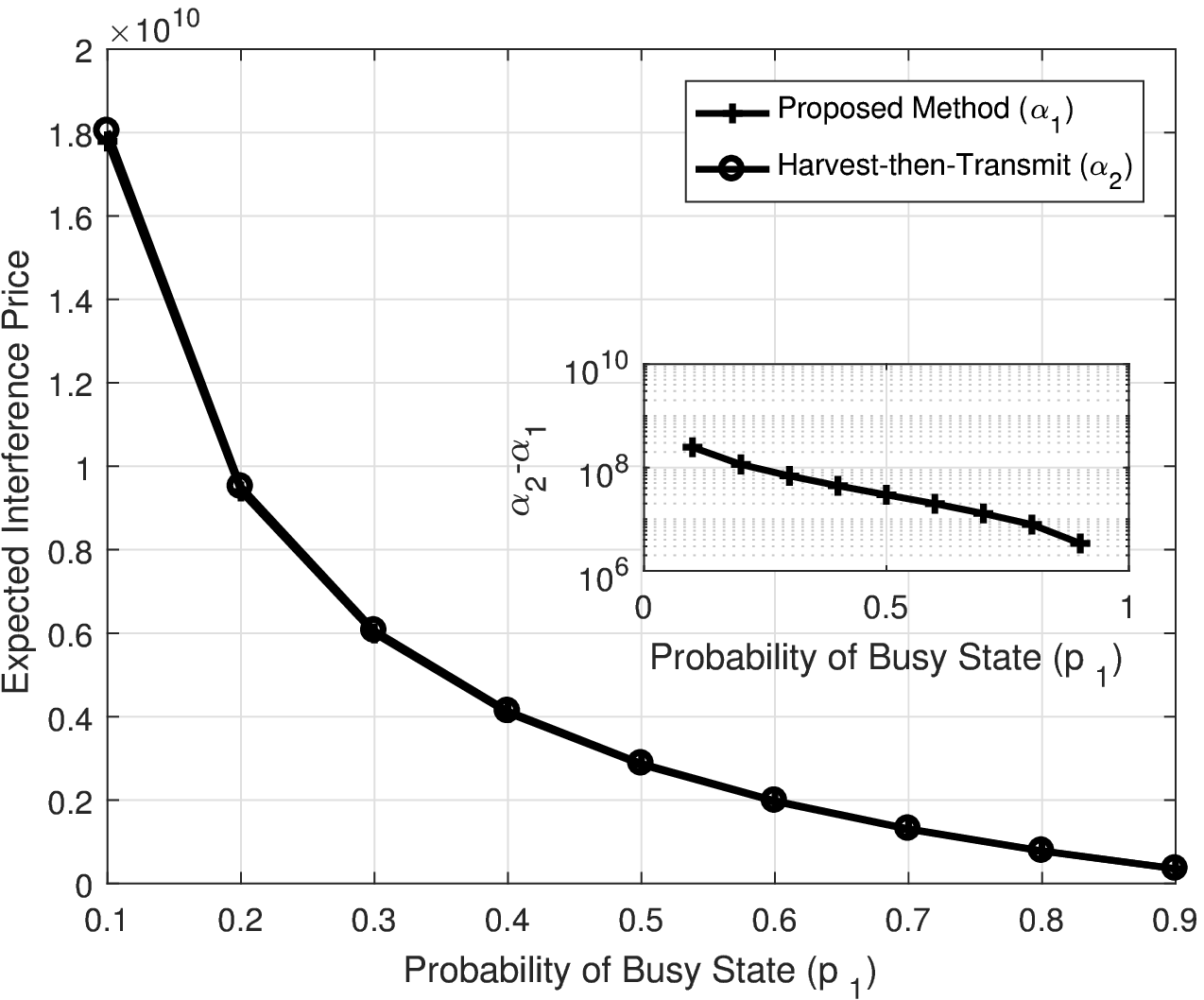}}\hspace{5mm}
\subfigure[]{\label{fig_probability_comparison_d}\includegraphics[width=.34\linewidth]{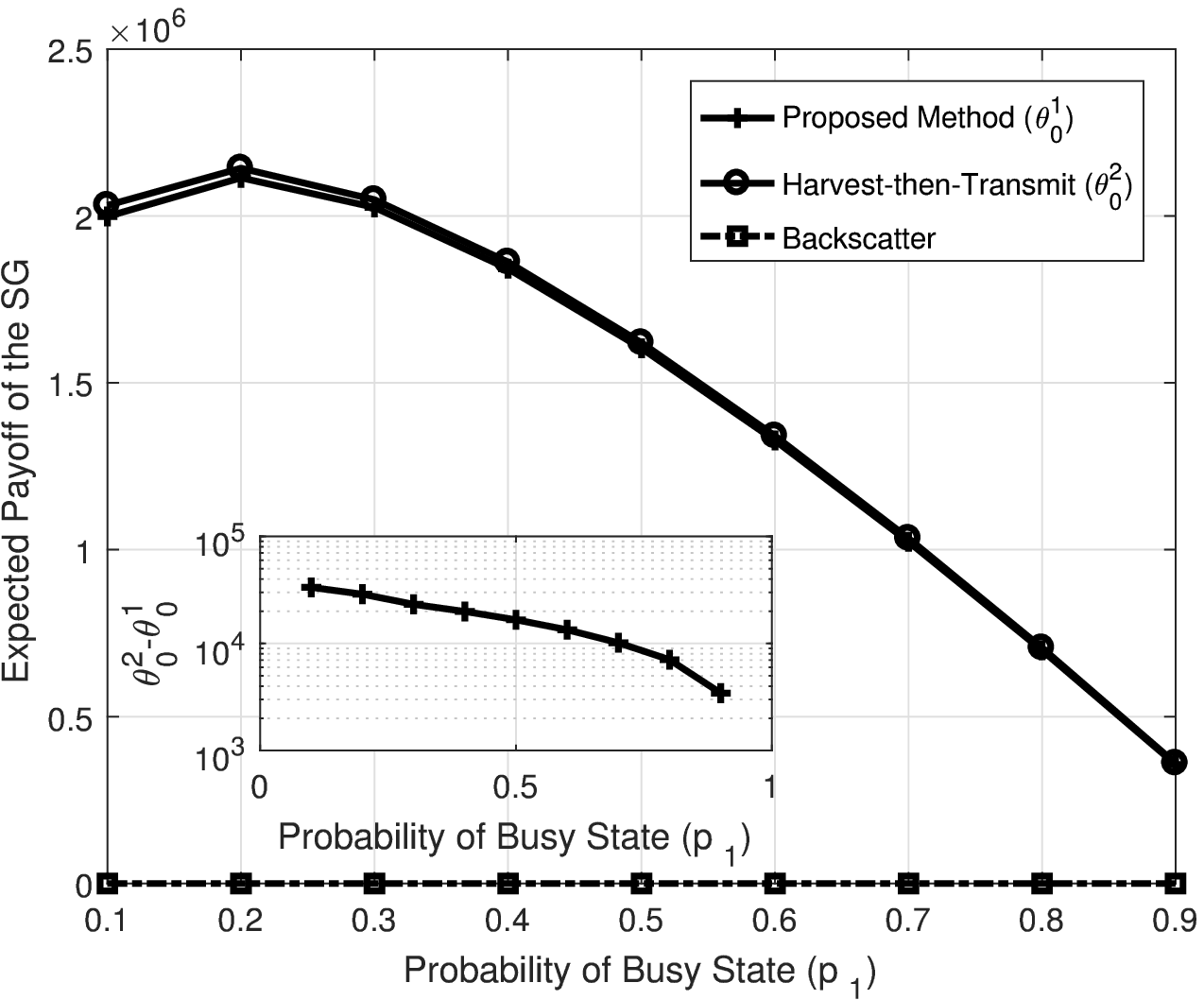}}\vspace*{-1mm}
\caption{Performance comparison between the proposed scheme and two reference schemes. (a) Expected average payoff of the STs at the SE vs. different $p_1$. (b) Expected average throughput of the STs at the SE vs. different $p_1$. (c) Expected interference price at the SE vs. different $p_1$. (d) Expected payoff of the SG at the SE vs. different $p_1$.}
\label{fig_probability_comparison}\vspace*{-3mm}
\end{figure*}
In Figure~\ref{fig_probability_comparison}, we investigate the impact of the probability of the busy state on the performance of the proposed transmission scheme. For the simulation, the number of the STs is fixed at 5. We note from Figures~\ref{fig_probability_comparison_a} and~\ref{fig_probability_comparison_b} that the performance of the backscatter-only scheme improves as the probability of the channel staying busy increases, while the performance of the harvest-then-transmit-only scheme becomes worse at the same time. We can further observe from Figure~\ref{fig_probability_comparison_b} that as the chance of direct transmission reduces with the increasing probability of the channel staying busy, with the proposed hybrid transmission policy, the STs are able to achieve a significantly higher throughput than that of using either of the two fixed-scheme transmission policies. Especially, the proposed hybrid scheme suffers from less severe performance deterioration than that of the harvest-then-transmit scheme. Again, as can be interpreted from Figures~\ref{fig_probability_comparison_c} and~\ref{fig_probability_comparison_d}, by adopting the proposed transmission scheme, the STs have an advantage in interference price negotiation with the SG over the harvest-then-transmit-only scheme. When the chance of transmission becomes smaller as the channel becomes busier, such an advantage will lead to a significant performance improvement at the SE.\vspace*{-3mm}

\section{Conclusion}
\label{sec_conclusion}
In this paper, we have studied the sensing-pricing-transmitting strategy adaptation problem in the RF-powered backscatter cognitive radio network. In the considered network, the secondary transmitters operate in a non-cooperative manner to compete for the time resource for their own transmission. The secondary gateway considers the condition of imperfect channel detection and employs a pricing mechanism to manage the interference from the secondary network to the primary users. Mathematically, we have modeled the strategy adaptation process between the secondary gateway and the secondary transmitters as a single-leader-multi-follower Stackelberg game, where the follower sub-game among the secondary transmitters is further modeled as a generalized Nash equilibrium problem with shared constraints. Based on our discoveries on the uniqueness of the generalized Nash equilibrium in the follower sub-game and the continuity of the leader's strategy-payoff mapping, we have proposed a directional ascent-based algorithm for the leader's strategy searching, where a distributed Nash equilibrium searching algorithm for the lower-level sub-game is nested therein. Both of the theoretical and numerical analysis have shown the convergence and efficiency of the proposed scheme.

\appendices
\section{Proof of Theorem \ref{thm_convex_game}}
\label{Proof_thm_convex_game}
The property in {\bf{P1}} of Theorem~\ref{thm_convex_game} is straightforward from the formulation of (\ref{eq:o_modifiedpt:local}). To prove {\bf{P2}}, we analyze the Hessian matrix $H_k = \nabla^2_{s^2_k}\theta_k(\mathbf{s}^{\textrm{ST}})$ for each player $k$ given a feasible adversaries' strategy $s^{\textrm{ST}}_{-k}$. Note that the value of $\theta_k(\mathbf{s}^{\textrm{ST}})$ is determined only by $s_k$, then, we have
 \begin{equation}
 \label{eq_hessian}
  H_k=
  \begin{bmatrix}
      \displaystyle\frac{\partial^2\theta_k}{{\partial(\tau^h_k})^2} & \displaystyle\frac{\partial^2\theta_k}{\partial\tau^h_k\partial\tau^t_k} &
      \displaystyle\frac{\partial^2\theta_k}{\partial\tau^h_k\partial\tau^b_k}\\

      \displaystyle\frac{\partial^2\theta_k}{{\partial\tau^t_k}{\partial\tau^h_k}} & \displaystyle\frac{\partial^2\theta_k}{\partial(\tau^t_k)^2} &
      \displaystyle\frac{\partial^2\theta_k}{\partial\tau^t_k\partial\tau^b_k}\\

      \displaystyle\frac{\partial^2\theta_k}{{\partial\tau^b_k}{\partial\tau^h_k}} & \displaystyle\frac{\partial^2\theta_k}{\partial\tau^b_k\partial\tau^t_k} &
      \displaystyle\frac{\partial^2\theta_k}{\partial(\tau^b_k)^2}
  \end{bmatrix}.
 \end{equation}
We note that $H_k$ is symmetric. Given a fixed pair of $(\tau^s,\epsilon)$, let us define
 \begin{equation}
  \label{eq_params_As}
  \left\{\begin{array}{ll}
  A^1_k = \nu p_0(1-p^f(\tau^s,\epsilon))\kappa_k W,\\
  A^2_k = \nu p_1(1-p^d(\tau^s,\epsilon))\kappa_k W,
  \end{array}\right.
 \end{equation}
and
\begin{equation}
  \label{eq_params_Gammas}
   \left\{\begin{array}{ll}
  \gamma^1_k = \displaystyle\frac{p_1p^d(\tau^s,\epsilon)\delta h_{k}h_k^{\textrm{PT}}P_{\textrm{PT}}}{\sigma^2_k},\\
  \gamma^2_k = \displaystyle\frac{p_1p^d(\tau^s,\epsilon)\delta h_{k}h_k^{\textrm{PT}}P_{\textrm{PT}}}{\sigma^2_k+h_k^{\textrm{PT}}P_{\textrm{PT}}}.\\
  \end{array}\right.
\end{equation}
Then, we have
\allowdisplaybreaks
\begin{align}
  \label{eq_hessian_1}
 &\frac{\partial^2\theta_k}{\partial(\tau^b_k)^2}=\frac{\partial^2\theta_k}{{\partial\tau^b_k}{\partial\tau^h_k}}=\frac{\partial^2\theta_k}{\partial\tau^b_k\partial\tau^t_k}=0,\\
 \label{eq_hessian_2}
 &\frac{\partial^2\theta_k}{{\partial(\tau^t_k})^2}=-\frac{A_k^1 (\gamma_k^1)^2(\tau_k^h)^2}{(1-P^c+\gamma_k^1\tau_k^h/\tau_k^t)^2(\tau_k^t)^3\ln2}-
 \frac{A_k^2 (\gamma_k^2)^2(\tau_k^h)^2}{(1-P^c+\gamma_k^2\tau_k^h/\tau_k^t)^2(\tau_k^t)^3\ln2},\\
 \label{eq_hessian_3}
 &\frac{\partial^2\theta_k}{{\partial\tau^h_k}{\partial\tau^t_k}}=\frac{\partial^2\theta_k}{{\partial\tau^t_k}{\partial\tau^h_k}}=\frac{A_k^1(\gamma_k^1)^2\tau_k^h}
 {(1-P^c+\gamma_k^1\tau_k^h/\tau_k^t)^2\tau_k^t\ln2}+\frac{A_k^2(\gamma_k^2)^2\tau_k^h}{(1-P^c+\gamma_k^2\tau_k^h/\tau_k^t)^2\tau_k^t\ln2},\\
 \label{eq_hessian_4}
 &\frac{\partial^2\theta_k}{{\partial(\tau^h_k})^2}=-\frac{A_k^1(\gamma_k^1)^2}{(1-P^c+\gamma_k^1\tau_k^h/\tau_k^t)^2\tau_k^t\ln2}-
 \frac{A_k^2(\gamma_k^2)^2}{(1-P^c+\gamma_k^2\tau_k^h/\tau_k^t)^2\tau_k^t\ln2}.
\end{align}
For any arbitrary real-valued non-zero vector $v_k=[v_k^1, v_k^2, v_k^3]^{\top}$, after substituting (\ref{eq_hessian_1})-(\ref{eq_hessian_4}) into (\ref{eq_hessian}), we obtain
\begin{equation}
  \label{eq_vector_product}
 v^{\top}_kH_kv_k\!=\!-\frac{\left(A_k^1(\gamma_k^1)^2(\gamma_k^2\tau_k^h\!+\!(1\!-\!P^c)\tau_k^t)^2+A_k^2(\gamma_k^2)^2(\gamma_k^1\tau_k^h\!+\!(1\!-\!P^c)\tau_k^t)^2\right)\left(\tau_k^tv_k^2\!-\!\tau_k^hv_k^1\right)^2}
 {\tau_k^t\left(\gamma_k^1\tau_k^h+(1\!-\!P^c)\tau_k^t\right)^2\left(\gamma_k^2\tau_k^h+(1\!-\!P^c)\tau_k^t\right)^2\ln2}\le0.
\end{equation}
Therefore, $H_k$ is negative semidefinite and $\theta_k(\mathbf{s}^{\textrm{ST}})$ is a concave function in $s_k$.\vspace*{-3.5mm}
\section{Proof of Theorem~\ref{thm_existence_NE_subgame}}
\label{app_unique_sub_game_GNE}
From Definition~\ref{def:Stackelberg_game}, we note that $\mathcal{S}^{\textrm{ST}}$ is a closed convex polytope and is therefore continuous. By {\bf{P2}} in Theorem~\ref{thm_convex_game}, $\theta_k(s_k, s^{\textrm{ST}}_{-k})$ is a $C^2$ concave function in $s_k$. We also note from (\ref{eq:o_modifiedpt:local}) that $\theta_k(s_k, s^{\textrm{ST}}_{-k})$ depends only on $s_k$. Therefore, we can construct the Jacobian of $F^f(\mathbf{s}^{\textrm{ST}})$, $\nabla_{\mathbf{s}^{\textrm{ST}}} F^f(\mathbf{s}^{\textrm{ST}})\!=\!-\left[\nabla^2_{s_k,s'_k}\theta_k(\mathbf{s})\right]_{k,k'=1}^{K}$ and obtain the following block diagonal matrix:
\begin{equation}
\label{eq_jacobian}
-\nabla_{\mathbf{s}^{\textrm{ST}}} F^f(\mathbf{s}^\textrm{ST})\!=\!
\begin{bmatrix}
 \nabla^2_{s^2_1}\theta_1(\mathbf{s}^\textrm{ST}) \!&\! \nabla^2_{s_1,s_2}\theta_1(\mathbf{s}^\textrm{ST}) \!&\! \cdots \!&\! \nabla^2_{s_1,s_K}\theta_1(\mathbf{s}^\textrm{ST})\\
 \nabla^2_{s_2,s_1}\theta_2(\mathbf{s}^\textrm{ST}) \!&\! \nabla^2_{s^2_2}\theta_2(\mathbf{s}^\textrm{ST}) \!&\! \cdots \!&\! \nabla^2_{s_2,s_K}\theta_2(\mathbf{s}^\textrm{ST})\\
  \vdots \!&\! \vdots \!&\! \ddots \!&\! \vdots\\
 \nabla^2_{s_K,s_1}\theta_K(\mathbf{s}^\textrm{ST}) \!&\! \nabla^2_{s_K,s_2}\theta_K(\mathbf{s}^\textrm{ST}) \!&\! \cdots \!&\! \nabla^2_{s^2_K}\theta_K(\mathbf{s}^\textrm{ST})\\
\end{bmatrix}
\!=\!
\begin{bmatrix}
                     H_1 & \mathbf{0} & \cdots & \mathbf{0}\\
                     \mathbf{0} & H_2 & \cdots & \mathbf{0}\\
                     \vdots & \vdots & \ddots & \vdots\\
                     \mathbf{0} & \mathbf{0} & \cdots & H_K
                     \end{bmatrix},
\end{equation}
where $H_k$ is given by (\ref{eq_hessian}) in Appendix~\ref{Proof_thm_convex_game}. With the same technique used in Appendix~\ref{Proof_thm_convex_game}, we can immediately verify that $\nabla_{\mathbf{s}^\textrm{ST}}F^f(\mathbf{s}^\textrm{ST})$ is positive semidefinite on $\mathcal{S}^\textrm{ST}$. Therefore, $F^f(\mathbf{s}^\textrm{ST})$ is a continuous monotone mapping on $\mathcal{S}^\textrm{ST}$. Then, according to (8) and (13) in \cite{5447064} (cf. Theorem~4.1 in~\cite{Facchinei2010}), the set of solutions to $\mathop{\textrm{VI}}(\mathcal{S}^\textrm{ST},F^f)$ is nonempty, closed and convex. By Lemma \ref{lemma_equivalence}, such a set of solution is at the same time the set of GNE for game $\mathcal{G}^f$. Then, a nonempty, closed and convex set of GNE exists in the follower sub-game $\mathcal{G}^f$.

Furthermore, we note from (\ref{eq:tx_rate_idle}) that $r^t_k(s_k,s_0)$ in the objective function of (\ref{eq:o_modifiedpt:local}) is a monotonic increasing function of $\tau^h_k$. Then, we can show by contradiction that for ST $k$'s local optimization problem in (\ref{eq:o_modifiedpt:local}), the equality in the constraint $\tau^b_k\!+\!\tau^h_k\!\le\!(T\!-\!\tau^s)$ is always reached at the sub-game NE strategy $s^*_k\!=\!(\tau^{h,*}_k,\tau^{t,*}_k,\tau^{b,*}_k)$, $\forall k\in{\mathcal{K}^{\textrm{ST}}}$. Otherwise, we can find a different strategy $s_k\!=\!(\tau^{h}_k,\tau^{t,*}_k,\tau^{b,*}_k)$, where $\tau^h\!=\!(T\!-\!\tau^s)\!-\!\tau^{b,*}_k\!>\!\tau^{h,*}_k$ such that $\theta_k(s_k;s_0)\!>\!\theta_k(s^*_k;s_0)$, which contradicts with (\ref{eq_equilibrium}) in Definition~\ref{def:Nash_equilibrium}. Thereby, the proof to Theorem~\ref{thm_existence_NE_subgame} is completed.

\section{Proof of Lemma~\ref{lemma_socail_optimal} and Theorem~\ref{thm_social_optimal}}
\label{app_socail_optimal}
\subsubsection{Proof of Lemma~\ref{lemma_socail_optimal}}
Following Lemma~\ref{lemma_potential}, by the definition of the potential game, we have $\nabla_{s_k} \phi(\mathbf{s}^{\textrm{ST}})\!=\!\nabla_{s_k} \theta_k(\mathbf{s}^{\textrm{ST}})$ (cf.~\cite{MONDERER1996124}). Then, with the extra condition $\tau^h_k\!=\!T\!-\!\tau^s\!-\!\tau^b_k$ given in Theorem~\ref{thm_existence_NE_subgame}, consider the following network payoff maximization problem based on the potential function given in (\ref{eq_potential_func}):
\begin{equation}
\label{eq:o_modifiedpt:num}
\begin{array}{ll}
\mathbf{s}^{\textrm{ST},*}=\arg\max\limits_{\mathbf{s}^{\textrm{ST}}=(s_k)_{k=1}^K} \!&\left(\phi(\mathbf{s}^{\textrm{ST}})\!=\!\sum\limits_{k=1}^K \Big(\nu\left(\tau^b_k r^b_k(\tau^s, \epsilon)\!+\!\tau^t_k r^t_k(s_k;\tau^s, \epsilon)\right)\!-\!\alpha p_1\left(1-p^d(\tau^s, \epsilon)\right)\tau^t_k\Big)\right),\\
\qquad\qquad\qquad\textrm{s.t.} & \sum\limits_{i=1}^{K}\tau^b_i\le(T-\tau^s), \sum\limits_{i=1}^{K}\tau^t_i\le(T-\tau^s),\\
 &\forall k\in{\mathcal{K}^{\textrm{ST}}}: \; \tau^t_k\ge0, \tau^b_k\ge0,\\
 &\forall k\in{\mathcal{K}^{\textrm{ST}}}: \; p_1(1-\overline{p}^m)\delta h_k^{\textrm{PT}}P_{{\textrm{PT}}}(T-\tau^s-\tau^b_k)-P^c\tau_k^t\ge0.
\end{array}
\end{equation}
By Theorem 2.2 in~\cite{MONDERER1996124}, the set of solutions to (\ref{eq:o_modifiedpt:num}) coincides with the set of NE in $\mathcal{G}^f$. By (\ref{eq_jacobian}),
$\phi(\mathbf{s}^{\textrm{ST}})$ is concave with respect to $\mathbf{s}^{\textrm{ST}}$ and (\ref{eq:o_modifiedpt:num}) is a concave programming problem. From (\ref{eq:o_modifiedpt:num}), we can obtain the following Lagrangian function:
\begin{equation}
\label{eq_Lagragian}
L(\mathbf{s}^{\textrm{ST}}, \pmb\lambda, \pmb\mu_1,\ldots,\pmb\mu_k) =\phi(\mathbf{s}^{{\textrm{ST}}})-G(\mathbf{s}^{{\textrm{ST}}})^{\top}\pmb\lambda-\sum_{k=1}^{K}Z(s_k)^{\top}\pmb\mu_k.
\end{equation}
Since $\forall k\!\in\!\mathcal{K}^{\textrm{ST}}$, $\nabla_{s_{j:j\ne k}} Z_k(s_k)\!=\!0$, from (\ref{eq_Lagragian}), the KKT condition to (\ref{eq:o_modifiedpt:num}) can be written as
\begin{align}
\label{eq_kkt_num_1}
&\forall k\in\mathcal{K}^{\textrm{ST}}:\;\nabla_{s_k}\phi(\mathbf{s}^{{\textrm{ST}}})-\left(\nabla_{s_k}G(\mathbf{s}^{{\textrm{ST}}})\right)^{\top}\pmb\lambda-\left(\nabla_{s_k}Z(s_k)\right)^{\top}\pmb\mu_k=0,\\
 \label{eq_kkt_num_2}
&\mathbf{0}\le\pmb\lambda\perp -G(\mathbf{s}^{{\textrm{ST}}})\ge 0,\\
 \label{eq_kkt_num_3}
&\forall k\in\mathcal{K}^{\textrm{ST}}:\;\mathbf{0}\le\pmb\mu_k\perp -Z_k({s}_k)\ge 0,
\end{align}
where $G(\mathbf{s}^{\textrm{ST}})$ and $Z_k(s_k)$ are given by (\ref{eq_joint_constraint}) and (\ref{eq_local_constraint}), respectively. By Theorem 2.2 in~\cite{MONDERER1996124}, from the solutions to (\ref{eq_kkt_num_1})-(\ref{eq_kkt_num_3}), i.e., $\left(\mathbf{s}^{\textrm{ST},*}, \pmb\lambda^*, \pmb\mu^*_1,\ldots,\pmb\mu^*_k\right)$, $\mathbf{s}^{\textrm{ST},*}$ forms a set of solutions which is equivalent to the set of NE in $\mathcal{G}^f$. Meanwhile, by Lemma~\ref{lemma_kkt_solution}, for each $\mathbf{s}^{\textrm{ST},*}$ there exists a set of multipliers $(\tilde{\pmb\lambda}_k, \tilde{\pmb\mu}_k)$, $\forall k\!\in\!\mathcal{K}^\textrm{ST}$, that forms a solution to the concatenated KKT conditions given by (\ref{eq_kkt_1})-(\ref{eq_kkt_3}). If we set $\tilde{\pmb\lambda}_k\!=\!\pmb\lambda^*$ and $\tilde{\pmb\mu}_k\!=\!\pmb\mu_k^*$, $\forall k\!\in\!\mathcal{K}^{\textrm{ST}}$, by comparing (\ref{eq_kkt_num_1})-(\ref{eq_kkt_num_3}) and (\ref{eq_kkt_1})-(\ref{eq_kkt_3}), we note that $\nabla_{s_k}\phi(\mathbf{s}^{\textrm{ST}})\!=\!\nabla_{s_k} \theta_k(\mathbf{s}^{\textrm{ST}})$, then, $\left(\mathbf{s}^{\textrm{ST},*}, \pmb\lambda^*_1=\cdots=\pmb\lambda^*_k=\pmb\lambda^*, \pmb\mu^*_1,\ldots,\pmb\mu^*_k\right)$ is exactly the solution to the concatenated KKT system given by (\ref{eq_kkt_num_1})-(\ref{eq_kkt_num_3}). Therefore, Lemma~\ref{lemma_socail_optimal} is proved.

\subsubsection{Proof of Theorem~\ref{thm_social_optimal}}
Due to the negative semi-definiteness of $-\nabla_{\mathbf{s}^{\textrm{ST}}} F^f(\mathbf{s}^\textrm{ST})$, which happens to be the Hessian matrix of $\phi(\mathbf{s}^{{\textrm{ST}}})$, we examine the Strong Sufficient Optimality Condition (SSOC) of second order for the Lagrangian given by (\ref{eq_Lagragian}) instead. For the specific convex optimization problem described by (\ref{eq_kkt_num_1})-(\ref{eq_kkt_num_3}), the SSOC is mathematically defined as follows:
\begin{Definition}[SSOC~\cite{dempe2002foundations}]
  \label{def_SSOC}
The SSOC is said to be satisfied at $\mathbf{s}=(s_0,\mathbf{s}^{\textrm{ST}})$, if for each set of multiplies $(\pmb\lambda, \pmb\mu_1,\ldots,\pmb\mu_K)$ that solves (\ref{eq_kkt_num_1})-(\ref{eq_kkt_num_3}) with $\mathbf{s}^{\textrm{ST}}$, for every non-zero vector $\mathbf{d}$ satisfying
\begin{equation}
  \label{eq_ssoc_inequal}
  \left\{\begin{array}{ll}
    \mathbf{d}^{\top}\nabla_{\mathbf{s}^{\textrm{ST}}} G^i(s_0, \mathbf{s}^{\textrm{ST}})=0,  \textrm{ if }  \lambda^i>0, \forall i\in\{1,2\},\\
    \mathbf{d}^{\top}\nabla_{\mathbf{s}^{\textrm{ST}}} Z^i_k(s_0, \mathbf{s}^{\textrm{ST}})=0, \textrm{ if } \mu^i_k>0, \forall k\in\mathcal{K}^{\textrm{ST}}, \forall i\in\{1,2, 3\}, \\
  \end{array}\right.
\end{equation}
we have $\mathbf{d}^{\top}\nabla^2_{\mathbf{s}^{\textrm{ST}},\mathbf{s}^{\textrm{ST}}}L(\mathbf{s}^{\textrm{ST}}, \pmb\lambda, \pmb\mu_1,\ldots,\pmb\mu_k)\mathbf{d}<0$, where $L(\mathbf{s}^{\textrm{ST}}, \pmb\lambda, \pmb\mu_1,\ldots,\pmb\mu_k)$ is defined in (\ref{eq_Lagragian}).
\end{Definition}

From the complementary conditions in (\ref{eq_kkt_num_2}) and (\ref{eq_kkt_num_3}), we know that the conditions $\lambda^i\!>\!0$ and $\mu_k^i\!>\!0$ hold only when $G^i\!=\!0$ and $Z^i_k\!=\!0$, respectively. Noting that $\nabla_{\mathbf{s}^{\textrm{ST}}} G^i(s_0, \mathbf{s}^{\textrm{ST}})$ and $\nabla_{\mathbf{s}^{\textrm{ST}}} Z^i_k(s_0, \mathbf{s}^{\textrm{ST}})$ are all constant vectors, we have $\nabla^2_{\mathbf{s}^{\textrm{ST}},\mathbf{s}^{\textrm{ST}}}L(\mathbf{s}^{\textrm{ST}}, \pmb\lambda, \pmb\mu_1,\ldots,\pmb\mu_k)\!=\!\nabla^2_{\mathbf{s}^{\textrm{ST}},\mathbf{s}^{\textrm{ST}}}\phi(\mathbf{s}^{\textrm{ST}})\!=\!-\nabla_{\mathbf{s}^{\textrm{ST}}}F^f(\mathbf{s}^\textrm{ST})$. Since (\ref{eq_jacobian}) defines a block diagonal matrix, by Definition~\ref{def_SSOC} and Lemma~\ref{lemma_socail_optimal}, it suffices to verify that $\forall k\in\mathcal{K}^{\textrm{ST}}$, $\tilde{\mathbf{d}}^{\top}\nabla^2_{s^2_k}\phi(\mathbf{s}^{\textrm{ST}})\tilde{\mathbf{d}}<0$, $\forall\tilde{\mathbf{d}}\ne0$ satisfying the following conditions:
\begin{equation}
  \label{eq_ssoc_inequal_local}
  \left\{\begin{array}{ll}
    \tilde{\mathbf{d}}^{\top}\nabla_{{s}_k} G^i(s_0, \mathbf{s}^{\textrm{ST}})=0, \textrm{ if } \lambda^i>0,\forall i\in\{1,2\},\\
    \tilde{\mathbf{d}}^{\top}\nabla_{{s}_k} Z^i_k(s_0, \mathbf{s}^{\textrm{ST}})=0, \textrm{ if } \mu^i_k>0, \forall i\in\{1,2,3\}.\\
  \end{array}\right.
\end{equation}

We note from (\ref{eq_joint_constraint}) and (\ref{eq_local_constraint}) that for ST $k$ with strategy $s_k\!=\!(\tau^t_k, \tau^b_k)$, $\nabla_{s_k} G^1(s_0, \mathbf{s}^{\textrm{ST}})\!=\![1, 0]^{\top}$, $\nabla_{s_k} G^2(s_0, \mathbf{s}^{\textrm{ST}})\!=\![0, 1]^{\top}$, $\nabla_{s_k} Z_k^1(s_0, \mathbf{s}^{\textrm{ST}})\!=\![-1, 0]^{\top}$, $\nabla_{s_k} Z_k^2(s_0, \mathbf{s}^{\textrm{ST}})\!=\![0, -1]^{\top}$ and $\nabla_{s_k} Z_k^3(s_0, \mathbf{s}^{\textrm{ST}})\!=\![P^c, p_1(1-\overline{p}^m)\delta h_k^{\textrm{PT}}P^{\textrm{PT}}]^{\top}$. By (\ref{eq_vector_product}), we have $\tilde{\mathbf{d}}^{\top}\nabla^2_{s^2_k}\phi(\mathbf{s}^{\textrm{ST}})\tilde{\mathbf{d}}=\tilde{\mathbf{d}}^{\top}\nabla^2_{s^2_k}\theta_k(s_k)\tilde{\mathbf{d}}\le0$ for any $\tilde{\mathbf{d}}$. From (\ref{eq_ssoc_inequal_local}), we note that to ensure $\tilde{\mathbf{d}}\ne0$, we are able to enumerate the following cases of non-zero multiplier combination without the need of knowing the set of $(s_k^*,\pmb\lambda^*,\pmb\mu^*_k)$:
\begin{itemize}
  \item [(i)] If $G^1(s_0, \mathbf{s}^{\textrm{ST}})=0$ or $Z^1(s_0, \mathbf{s}^{\textrm{ST}})=0$, and at the same time $G^2(s_0, \mathbf{s}^{\textrm{ST}})<0$, $Z^2(s_0, \mathbf{s}^{\textrm{ST}})<0$ and $Z^3(s_0, \mathbf{s}^{\textrm{ST}})<0$, we have $\tilde{\mathbf{d}}\!\in\!\{{\mathbf{d}}\!=\![d_1,0]^{\top}\!:\!d_1\!\in\!\mathbb{R},d_1\!\ne\!0\}$.
  \item [(ii)] if $G^2(s_0, \mathbf{s}^{\textrm{ST}})=0$ or $Z^2(s_0, \mathbf{s}^{\textrm{ST}})=0$, and at the same time $G^1(s_0, \mathbf{s}^{\textrm{ST}})<0$,
  $Z^1(s_0, \mathbf{s}^{\textrm{ST}})<0$ and $Z^3(s_0, \mathbf{s}^{\textrm{ST}})<0$, we have $\tilde{\mathbf{d}}\!\in\!\{{\mathbf{d}}\!=\![0, d_1]^{\top}\!:\!d_1\!\in\!\mathbb{R},d_1\!\ne\!0\}$.
  \item [(iii)] If $Z^3(s_0, \mathbf{s}^{\textrm{ST}})=0$, and at the same time $G^1(s_0, \mathbf{s}^{\textrm{ST}})<0$, $G^2(s_0, \mathbf{s}^{\textrm{ST}})<0$, $Z^1(s_0, \mathbf{s}^{\textrm{ST}})<0$ and $Z^2(s_0, \mathbf{s}^{\textrm{ST}})<0$, we have $\tilde{\mathbf{d}}\!\in\!\{{\mathbf{d}}\!=\![d_1, -\frac{P^c}{p_1(1-\tilde{p}^m)\delta h_k^{\textrm{PT}}P^{\textrm{PT}}}d_1]^{\top}\!:\!d_1\!\in\!\mathbb{R},d_1\!\ne\!0\}$.
\end{itemize}
From (\ref{eq_vector_product}), we can derive the condition of $\tilde{\mathbf{d}}^{\top}\nabla^2_{s^2_k}\theta_k(s_k)\tilde{\mathbf{d}}=0$ for ST $k$ in cases (i), (ii) and (iii) as $\tau^t_k=0$, $\tau^b_k=T-\tau^s$ and $\tau^b_k-\frac{p_1(1-\tilde{p}^m)\delta h_k^{\textrm{PT}}P^{\textrm{PT}}}{P^c}\tau^t_k=T-\tau^s$, respectively. Obviously, the equality condition for case (i) is satisfied only in the extreme scenario of $\tau^t_k=0, \forall k\in\mathcal{K}$, i.e., when no ST transmits at all. The equality conditions for cases (ii) and (iii) cannot be satisfied for every $k\in\mathcal{K}^{\textrm{ST}}$ at the same time.
Therefore, the SSOC is satisfied by the convex optimization problem described in (\ref{eq_kkt_num_1})-(\ref{eq_kkt_num_3}). This indicates that the optimal solution to (\ref{eq_kkt_num_1})-(\ref{eq_kkt_num_3}) is strongly stable (cf. Theorem 4.4 in~\cite{dempe2002foundations}), and the uniqueness of the solution is guaranteed. Then, Theorem~\ref{thm_social_optimal} immediately follows Lemma~\ref{lemma_socail_optimal}.\vspace*{-3.5mm}

\section{Proof of Theorem~\ref{thm_existence_GNE}}
\label{app_proof_SE_existence}
It is straightforward that $\theta_0(s_0, \mathbf{s}^{\textrm{ST}})$ in (\ref{eq_time_revenue_reformulated}) is continuous in $s_0$ and $\mathbf{s}^{\textrm{ST}}$, respectively, and coercive in the interference price $\alpha$. Meanwhile, $\mathcal{S}_0$ defined by (\ref{eq_time_revenue_reformulated_1})-(\ref{eq_time_revenue_reformulated_2}) is closed and continuous. Then, by the Weierstrass Theorem~\cite{sundaram1996first}, it is sufficient to prove that (a) $\Omega(s_0, \mathbf{s}^{\textrm{ST}})=\mathcal{S}_0\cap\mathop{\textrm{gph}}{\mathcal{E}(s_0)}$ is non-empty and closed, and (b) with the implicit function $\mathbf{s}^{\textrm{ST}}\!=\!\mathcal{E}(s_0)$, $\theta_0(s_0, \mathcal{E}(s_0))$ is continuous in $s_0$. Since the Slater's condition is satisfied for the constraints given by $G(\mathbf{s}^{\textrm{ST}})$ and $Z_k(s_k)$, $\forall k\!\in\!\mathcal{K}^{\textrm{ST}}$, by Proposition 3.2.7 in~\cite{Facchinei2003}, the Mangasarian-Fromovitz Constraint Qualification (MFCQ) (see~\cite{Facchinei2003,dempe2002foundations} for the definition) is satisfied at all feasible points $(s_0, \mathbf{s}^{\textrm{ST}})$ for the potential function-based lower-level optimization problem in (\ref{eq:o_modifiedpt:num}). Moreover, according to Appendix~\ref{app_socail_optimal}, the SSOC is also satisfied by the problem in (\ref{eq:o_modifiedpt:num}). Therefore, by Theorem 4.10 in~\cite{dempe2002foundations}, it is sufficient to prove that the following Constant Rank Constraint Qualification (CRCQ) is satisfied at any feasible point $\theta_0(s_0, \mathbf{s}^{\textrm{ST}})$.
\begin{Definition}[CRCQ~\cite{dempe2002foundations}]
  \label{def_CRCQ}
The CRCQ is satisfied for problem (\ref{eq:o_modifiedpt:num}) at a point $(s_0, \mathbf{s}^{\textrm{ST}})$, if there exists an open neighborhood $\mathcal{N}_{\rho}(s_0, \mathbf{s}^{\textrm{ST}})\!=\!\{(x,y): \Vert (x, y)-(s_0, \mathbf{s}^{\textrm{ST}})\Vert\!<\!\rho\}$ such that, for each subset of the active constraints defined as follows:
\begin{equation}
  \label{eq_active}
  I\subseteq I_0(s_0, \mathbf{s}^{\textrm{ST}})=\left\{i\!\in\!\{1,2\}, k\!\in\!\mathcal{K}^{\textrm{ST}}, k_j\!\in\!\{1,2,3\}: G^i(s_0, \mathbf{s}^{\textrm{ST}})=0, Z_k^{k_j}(s_0, \mathbf{s}^{\textrm{ST}})=0\right\}\nonumber,
\end{equation}
the family of gradient vectors $\{\nabla_{\mathbf{s}^{\textrm{ST}}}G^i(s_0, \mathbf{s}^{\textrm{ST}}), \nabla_{\mathbf{s}^{\textrm{ST}}}Z^{k_j}_k(s_0, \mathbf{s}^{\textrm{ST}}):(i,k,k_j)\!\in\!I\}$ has the same rank $\forall (x,y)\in\mathcal{N}_{\rho}(s_0, \mathbf{s}^{\textrm{ST}})$.
\end{Definition}
According to our discussion in Appendix~\ref{app_socail_optimal}, $\nabla_{\mathbf{s}^{\textrm{ST}}} G^i(s_0, \mathbf{s}^{\textrm{ST}})$ and $\nabla_{\mathbf{s}^{\textrm{ST}}} Z^j_k(s_0, \mathbf{s}^{\textrm{ST}})$ are all constant vectors $\forall i\!=\!1,2, j\!=\!1,2,3, k\!\in\!\mathcal{K}^{\textrm{ST}}$. Therefore, as long as the subset of active constraints $I$ is determined for $(s_0, \mathbf{s}^{\textrm{ST}})$, the family of gradient vectors $\{\nabla_{\mathbf{s}^{\textrm{ST}}}G^i(s_0, \mathbf{s}^{\textrm{ST}}), \nabla_{\mathbf{s}^{\textrm{ST}}}Z^{k_j}_k(s_0, \mathbf{s}^{\textrm{ST}}):(i,k,k_j)\!\in\!I\}$ are identical for all $\forall (x,y)\!\in\!\mathcal{N}_{\rho}(s_0, \mathbf{s}^{\textrm{ST}})$. Therefore, the CRCQ is satisfied everywhere in the problem given by (\ref{eq:o_modifiedpt:num}). By Theorem 4.10 in~\cite{dempe2002foundations}, $\mathbf{s}^{\textrm{ST}}\!=\!\mathcal{E}(s_0)$ is a piecewise continuously differentiable ($PC^1$) function and $\theta_0(s_0, \mathcal{E}(s_0))$ is therefore continuous in $s_0$. Furthermore, with the closed set $\mathcal{S}_0$, by the well-known Closed Graph Theorem, $\mathcal{E}(s_0)$ being continuous implies that $\mathop{\textrm{gph}}{\mathcal{E}(s_0)}$ is closed. Therefore, $\Omega(s_0, \mathbf{s}^{\textrm{ST}})$ is non-empty and closed, and the optimization problem in (\ref{eq_time_revenue_reformulated}) admits a globally optimal solution, namely an SE.\vspace*{-3.5mm}

\section{Proof of Theorem~\ref{thm_follower_convergence} and Corollary~\ref{cor_jacobian_convergence}}
\label{app_proof_uniqueness}
\subsubsection{Proof of Theorem~\ref{thm_follower_convergence}}
We first prove that if a sequence $\{\mathbf{s}^{\textrm{ST}}(t)\}_{t=0}^{\infty}$ generated by Algorithm~\ref{alg_heuristic} converges, then, that sequence converges to a GNE of $\mathcal{G}^f$. By Lemma~\ref{lemma_potential} and from (\ref{eq_best_response_update}), $\forall t, k$,
\begin{equation}
\label{proof_1_1}
  \begin{array}{ll}
  \phi(s_k(t+1), s^{\textrm{ST}}_{-k}(t))-\phi(s_k(t), s^{\textrm{ST}}_{-k}(t))\\
   =\theta_k(s_k(t+1), s^{\textrm{ST}}_{-k}(t))-\theta_k(s_k(t), s^{\textrm{ST}}_{-k}(t))\ge\displaystyle\frac{1}{2}\Vert s_k(t+1)-s_k(t)\Vert^2\ge0.
  \end{array}
\end{equation}
By {\bf{P1}} in Theorem~\ref{thm_convex_game}, given $k\!\in\!\mathcal{K}^{\textrm{ST}}$, $\forall s^{\textrm{ST}}_{-k}(t)$, we can always obtain a solution $s_k(t+1)$ to (\ref{eq_best_response_update}). From (\ref{eq_set_adversary}), we have $s^{\textrm{ST}}_{-(k+1)}(t)\!=\!(s_1(t\!+\!1),\ldots, s_k(t\!+\!1), s_{k+2}(t),\ldots, s_K(t))$, therefore, $(s_{k+1}(t), s^{\textrm{ST}}_{-(k+1)}(t))\!=\!(s_k(t\!+\!1), s^{\textrm{ST}}_{-k}(t))$. Let $\mathbf{s}^{\textrm{ST}}_k(t)$ denote the joint strategy updated in the inner iteration of Algorithm \ref{alg_heuristic}, $\mathbf{s}^{\textrm{ST}}_k(t)\!=\!(s_1(t\!+\!1),\ldots, s_k(t\!+\!1), s_{k+1}(t),\ldots, s_K(t))$. By (\ref{proof_1_1}) we have
\begin{equation}
 \label{proof_1_2}
 \phi(\mathbf{s}^{\textrm{ST}}(t+1))=\phi(\mathbf{s}^{\textrm{ST}}_K(t))\ge\phi(\mathbf{s}^{\textrm{ST}}_{K-1}(t))\ge\cdots\ge \phi(\mathbf{s}^{\textrm{ST}}_k(t))\ge\cdots\ge\phi(\mathbf{s}_0^{\textrm{ST}}(t))=\phi(\mathbf{s}^{\textrm{ST}}(t)).
\end{equation}

Assume that $\{\mathbf{s}^{\textrm{ST}}(t)\}_{t=0}^{\infty}$ is an infinite joint strategy sequence generated by Algorithm \ref{alg_heuristic} such that $\exists\overline{\mathbf{s}}^{\textrm{ST}}$, $\lim\limits_{t\rightarrow\infty}\mathbf{s}^{\textrm{ST}}(t)\!=\!\overline{\mathbf{s}}^{\textrm{ST}}$. Then, by the continuity of the potential function $\phi(\mathbf{s}^{\textrm{ST}})$, we know that the corresponding sequence $\{\phi(\mathbf{s}^{\textrm{ST}}(t))\}_{t=0}^{\infty}$ also converges, namely, $\lim\limits_{t\rightarrow\infty}\phi(\mathbf{s}^{\textrm{ST}}(t))\!=\!\overline\phi$. Therefore, for the sequence $\{\mathbf{s}^{\textrm{ST}}_k(t)\}$ generated by Algorithm~\ref{alg_heuristic}, based on (\ref{proof_1_2}) we have
\begin{equation}
 \label{eq_proof_1_3}
 \lim_{t\rightarrow\infty}\phi(\mathbf{s}^{\textrm{ST}}_k(t))=\overline{\phi}, \quad\forall k.
\end{equation}
Consider the sequence $\{s_k(t)\}_{t=0}^{\infty}$ extracted from $\{\mathbf{s}^{\textrm{ST}}(t)\}_{t=0}^{\infty}$, by (\ref{proof_1_1}) and (\ref{eq_proof_1_3}),
\begin{equation}
 \label{eq_proof_1_4}
 \lim_{t\rightarrow\infty}\left(\theta_k(s_k(t+1), s^{\textrm{ST}}_{-k}(t))-\theta_k(s_k(t), s^{\textrm{ST}}_{-k}(t))\right) = 0.
\end{equation}
Then, together with (\ref{proof_1_1}), we deduce that the following holds
\begin{equation}
 \label{eq_proof_1_5}
 \lim_{t\rightarrow\infty}\Vert s_k(t+1)-s_k(t)\Vert=0, \quad\forall k,
\end{equation}
or equivalently, $\lim\limits_{t\rightarrow\infty}\mathbf{s}^{\textrm{ST}}_k(t)=\overline{\mathbf{s}}^{\textrm{ST}}$, $\forall k$.

In order to prove that $\overline{\mathbf{s}}^{\textrm{ST}}$ is a GNE, we assume that there exists at least one local strategy vector for some ST $k$, $x_k\in\mathcal{S}(\overline{s}^{\textrm{ST}}_{-k})$, such that $\theta_k(x_k, \overline{s}^{\textrm{ST}}_{-k})>\theta_k(\overline{s}_k, \overline{s}^{\textrm{ST}}_{-k})$. Then, by the elementary properties of a concave function, we have
\begin{equation}
 \label{eq_proof_1_6}
 0<\theta_k(x_k, \overline{s}^{\textrm{ST}}_{-k}) - \theta_k(\overline{s}_k, \overline{s}^{\textrm{ST}}_{-k})\le \nabla^T_{s_k}\theta_k(\overline{\mathbf{s}})(x_k-\overline{s}_k).
\end{equation}
Let $\Theta_k(s_k, s^{\textrm{ST}}_{-k}, z)=\theta_k(s_k, s^{\textrm{ST}}_{-k})-\frac{1}{2}\Vert s_k-z \Vert^2$. It is straightforward to check the Hessian matrix of $\Theta_k(s_k, s^{\textrm{ST}}_{-k}, z)$ based on (\ref{eq_jacobian}) and show that the regularization term turns $\Theta_k(s_k, s^{\textrm{ST}}_{-k}, z)$ into a strongly concave function of $s_k$. Therefore, the uniqueness of the maximum solution to $s_k$ is guaranteed. Then, consider the directional derivative of $\Theta_k(s_k, s^{\textrm{ST}}_{-k}, z)$ in the direction $v=(x_k-\overline{s}_k)$ at a snapshot of the updating process in Algorithm \ref{alg_heuristic}:
\begin{equation}
 \label{eq_proof_1_7}
 \begin{array}{ll}
   \Theta'_k(s_k(t+1), s^{\textrm{ST}}_{-k}(t), s_k(t); (x_k-{s}_k(t+1)))\\
   =\theta'_k(s_k(t+1), s^{\textrm{ST}}_{-k}(t); (x_k-{s}_k(t+1))) + (s_k(t+1)-s_k(t))^{\top}(x_k-s_k(t+1))\\
   =\nabla^{\top}_{s_k}\theta_k(s_k(t+1), s^{\textrm{ST}}_{-k}(t))(x_k-{s}_k(t+1)) + (s_k(t+1)-s_k(t))^{\top}(x_k-s_k(t+1)).
 \end{array}
\end{equation}
By (\ref{eq_proof_1_5}) and the compactness of $\mathcal{S}_k$ (i.e., {\bf{P1}} in Theorem~\ref{thm_convex_game}), we have $\lim\limits_{t\rightarrow\infty}(s_k(t+1)-s_k(t))^{\top}
(x_k-s_k(t+1))=0$. According to the optimality condition for concave problems, from (\ref{eq_proof_1_7}) we obtain
\begin{equation}
 \label{eq_proof_1_8}
 \lim_{t\rightarrow\infty}\Theta'_k(s_k(t+1), s^{\textrm{ST}}_{-k}(t), s_k(t); (x_k-{s}_k(t+1)))
 =\nabla^{\top}_{s_k}\theta_k(\overline{\mathbf{s}}^{\textrm{ST}}_k)(x_k-\overline{s}_k)\le 0,
\end{equation}
if $s_k(t\!+\!1)$ is the solution to (\ref{eq_best_response_update}). Since (\ref{eq_proof_1_8}) contradicts with (\ref{eq_proof_1_6}), by the definition of an NE, we are ready to conclude that, if the sequence $\{\mathbf{s}^{\textrm{ST}}(t)\}_{t=0}^{\infty}$ converges, it converges to a GNE.

Now, it suffices to prove that the iteration given by Algorithm~\ref{alg_heuristic} is a contractive mapping~\cite{Facchinei2003} in order to show that Algorithm~\ref{alg_heuristic} converges to a GNE. From $\Theta(\mathbf{s}^{\textrm{ST}}, z)$ and (\ref{eq_best_response_update}), we can construct for each iteration an equivalent QVI problem $\mathop{\textrm{VI}}(\mathcal{S}^{\textrm{ST}},\tilde{F}^f(\mathbf{s};\mathbf{z}))$ with $\tilde{F}^f(\mathbf{s};\mathbf{z})=-(\nabla_{s_k}\theta_k({s}_k(\mathbf{z}), z_{-k})-(s_k-z_k))_{k=1}^{K}$. In order to construct the contraction map, we consider two strategy vectors $\mathbf{y}$ and $\mathbf{z}$, $\forall \mathbf{y}\ne\mathbf{z}, \mathbf{y},\mathbf{z}\in\mathcal{S}^{\textrm{ST}}$. Assume that $\mathbf{s}(\mathbf{y})$ and $\mathbf{s}(\mathbf{z})$ are the unique solutions to the strongly monotone QVI problems, $\tilde{F}^f(\mathbf{s};\mathbf{y})$ and $\tilde{F}^f(\mathbf{s};\mathbf{z})$, respectively. Then, by Definition \ref{def_VI} we have, $\forall k$,
\begin{align}
  \label{eq_proof_1_9}
   (x_k-s_k(\mathbf{y}))^{\top}\left(\nabla_{s_k}\theta_k({s}_k(\mathbf{y}), y_{-k})-(s_k(\mathbf{y})-y_k)\right)\le0, \forall x_k\in\mathcal{S}^{\textrm{ST}}_k(y_{-k}),\\
  \label{eq_proof_1_9_1}
   (x_k-s_k(\mathbf{z}))^{\top}\left(\nabla_{s_k}\theta_k({s}_k(\mathbf{z}), z_{-k})-(s_k(\mathbf{z})-z_k)\right)\le0, \forall x_k\in\mathcal{S}^{\textrm{ST}}_k(z_{-k}).
\end{align}
Let $x_k\!=\!s_k(\mathbf{z})$ in (\ref{eq_proof_1_9}) and $x_k\!=\!s_k(\mathbf{y})$ in (\ref{eq_proof_1_9_1}). By adding (\ref{eq_proof_1_9}) and (\ref{eq_proof_1_9_1}) together we obtain
\begin{equation}
 \label{eq_proof_1_10}
 \begin{array}{ll}
  (s_k(\mathbf{z})\!-\!s_k(\mathbf{y}))^{\top}\left(\nabla_{s_k}\theta_k({s}_k(\mathbf{y}), y_{-k})\!-\!s_k(\mathbf{y})\!+\!y_k\right)
  \!+\!(s_k(\mathbf{y})\!-\!s_k(\mathbf{z}))^{\top}\left(\nabla_{s_k}\theta_k({s}_k(\mathbf{z}), z_{-k})\!-\!s_k(\mathbf{z})\!+\!z_k\right)\\
  =(s_k(\mathbf{z})\!-\!s_k(\mathbf{y}))^{\top}\left(\nabla_{s_k}\theta_k({s}_k(\mathbf{y}), y_{-k})\!-\!\nabla_{s_k}\theta_k({s}_k(\mathbf{z}), z_{-k})\right)\!+\!
  (s_k(\mathbf{z})\!-\!s_k(\mathbf{y}))^{\top}(s_k(\mathbf{z})\!-\!s_k(\mathbf{y}))\\
  -(s_k(\mathbf{z})\!-\!s_k(\mathbf{y}))^{\top}(z_k-y_k)\le 0.
 \end{array}
\end{equation}
By the mean-value theorem, there exists a $0<\eta_k<1$ with $\tilde{x}_k=\eta_ks_k(\mathbf{z}) + (1-\eta_k)s_k(\mathbf{y})$, such that $\nabla_{s_k}\theta_k({s}_k(\mathbf{y}), y_{-k})\!-\!\nabla_{s_k}\theta_k({s}_k(\mathbf{z}), z_{-k})=\sum_{j\in\mathcal{K}^f}\nabla^2_{s_ks_j}\theta_k(\tilde{\mathbf{x}})(s_j(\mathbf{y})-s_j(\mathbf{z}))$. We note from (\ref{eq_jacobian}) that for any $j\ne k$, $\nabla^2_{s_ks_j}\theta_k(\tilde{\mathbf{x}})$ is a zero matrix. Then, from (\ref{eq_proof_1_10}) we have
\begin{eqnarray}
 \label{eq_proof_1_11}
 \begin{array}{ll}
  (s_k(\mathbf{z})\!-\!s_k(\mathbf{y}))^{\top}\nabla^2_{s^2_k}\theta_k(\tilde{\mathbf{x}})(s_k(\mathbf{y})\!-\!s_k(\mathbf{z}))\!+\!\Vert s_k(\mathbf{z})\!-\!s_k(\mathbf{y})\Vert^2
  \!-\!(s_k(\mathbf{z})\!-\!s_k(\mathbf{y}))^{\top}(z_k-y_k)\\
  =(s_k(\mathbf{z})\!-\!s_k(\mathbf{y}))^{\top}\left(-\nabla^2_{s^2_k}\theta_k(\tilde{\mathbf{x}})+I\right)(s_k(\mathbf{z})\!-\!s_k(\mathbf{y}))
  \!-\!(s_k(\mathbf{z})\!-\!s_k(\mathbf{y}))^{\top}(z_k-y_k)\!\le\!0.
 \end{array}
\end{eqnarray}
From (\ref{eq_proof_1_11}), we obtain
\begin{eqnarray}
 \label{eq_proof_1_12}
  \left\Vert-\nabla^2_{s^2_k}\theta_k(\tilde{\mathbf{x}})+I\right\Vert\left\Vert s_k(\mathbf{y})\!-\!s_k(\mathbf{z})\right\Vert\le\left\Vert z_k-y_k\right\Vert,
\end{eqnarray}
By {\bf{P2}} in Theorem~\ref{thm_convex_game}, $-\nabla^2_{s^2_k}\theta_k(\tilde{\mathbf{x}})$ is positive semidefinite, and all the eigenvalues of $-\nabla^2_{s^2_k}\theta_k(\tilde{\mathbf{x}})$ are non-negative. Furthermore, all the eigenvalues of matrix $\left(-\nabla^2_{s^2_k}\theta_k(\tilde{\mathbf{x}})+I\right)$ are no less than 1. By the elementary property of the eigenvalues and the spectral norm of a symmetric matrix, we have $\left\Vert-\nabla^2_{s^2_k}\theta_k(\tilde{\mathbf{x}})+I\right\Vert\ge1$, where the equality only happens when $\tilde{x}_{k,1}=\tilde{x}_{k,2}=0$. Therefore, the asynchronous proximal-response given by Algorithm \ref{alg_heuristic} is a contractive mapping, which completes the proof of Theorem \ref{thm_follower_convergence}.

\subsubsection{Proof of Corollary \ref{cor_jacobian_convergence}}
\label{subsec_proof_jacobian_convergence}
The proof of Corollary \ref{cor_jacobian_convergence} immediately follows our discussion on (\ref{eq_proof_1_12}). From (\ref{eq_proof_1_12}), we can construct the following vector inequality:
\begin{eqnarray}
 \label{eq_proof_1_13}
 \left(\begin{matrix}
  \left\Vert s_1(\mathbf{y})\!-\!s_1(\mathbf{z})\right\Vert\\
  \vdots\\
  \left\Vert s_K(\mathbf{y})\!-\!s_K(\mathbf{z})\right\Vert
 \end{matrix}\right)
 \le
 \left(\begin{matrix}
  {\left\Vert-\nabla^2_{s^2_1}\theta_1(\tilde{\mathbf{x}})+I\right\Vert}^{-1} & \ldots & 0\\
  0 & \ddots & 0\\
  0 & \ldots & {\left\Vert-\nabla^2_{s^2_K}\theta_K(\tilde{\mathbf{x}})+I\right\Vert}^{-1}
 \end{matrix}\right)
\left(\begin{matrix}
  \left\Vert z_1-y_1\right\Vert\\
  \vdots\\
  \left\Vert z_K-y_K\right\Vert
 \end{matrix}\right).
\end{eqnarray}
Recall that for the matrix norm if the component-wise inequality $A\ge B\ge0$ holds for two non-negative matrices $A$ and $B$, then $\Vert A\Vert\ge\Vert B\Vert$. Then, from (\ref{eq_proof_1_13}) the following holds
\begin{equation}
 \label{eq_proof_1_14}
 \begin{array}{ll}
 \Vert \mathbf{s}(\mathbf{y})\!-\!\mathbf{s}(\mathbf{z})\Vert\!=\!
 \left\Vert\left(\begin{matrix}
  \left\Vert s_1(\mathbf{y})\!-\!s_1(\mathbf{z})\right\Vert\\
  \vdots\\
  \left\Vert s_K(\mathbf{y})\!-\!s_K(\mathbf{z})\right\Vert
 \end{matrix}\right)\right\Vert
 \!\le\!
 \left\Vert\mathop{\textrm{Diag}}\left({\left\Vert-\nabla^2_{s^2_k}\theta_k(\tilde{\mathbf{x}})\!+\!I\right\Vert}^{-1}\right)_{k=1}^K\right\Vert
 \left\Vert\left(\begin{matrix}
  \left\Vert z_1\!-\!y_1\right\Vert\\
  \vdots\\
  \left\Vert z_K\!-\!y_K\right\Vert
 \end{matrix}\right)\right\Vert\\
 =\left\Vert\mathop{\textrm{Diag}}\left({\left\Vert-\nabla^2_{s^2_k}\theta_k(\tilde{\mathbf{x}})+I\right\Vert}^{-1}\right)_{k=1}^K\right\Vert
 \left\Vert \mathbf{y}-\mathbf{z}\right\Vert.
 \end{array}
\end{equation}
Since $\mathbf{s}(\mathbf{y})$ and $\mathbf{s}(\mathbf{z})$ are obtained based on simultaneous best-response updating from the strategies $\mathbf{y}$ and $\mathbf{z}$, we know that Algorithm \ref{alg_jacobian} is a contraction of best response and correspondingly with any initialization it converges to an NE. This completes the proof of Corollary \ref{cor_jacobian_convergence}.\vspace*{-1mm}

\linespread{1.27}
\bibliographystyle{IEEEtran}
\bibliography{Reference}
%
\end{document}